\documentclass{article}
\usepackage[main, final]{neurips_2026}

\usepackage[utf8]{inputenc} 
\usepackage[T1]{fontenc}    
\usepackage{hyperref}       
\usepackage{url}            
\usepackage{booktabs}       
\usepackage{amsfonts}       
\usepackage{nicefrac}       
\usepackage{microtype}      
\usepackage{xcolor}         

\usepackage{amsmath}
\usepackage{amssymb}
\usepackage{mathtools}
\usepackage{amsthm}
\usepackage{graphicx}
\usepackage{subcaption}
\usepackage{wrapfig}
\usepackage{multirow}
\usepackage{enumitem}

\theoremstyle{plain}
\newtheorem{theorem}{Theorem}[section]
\newtheorem{proposition}[theorem]{Proposition}

\theoremstyle{definition}

\theoremstyle{remark}

\title{Contrastive Retrieval Heads for Improved Attention-Based Reranking}

\author{%
Linh Tran$^{1, 2, *}$ \hspace{0.3em} Yulong Li$^{1}$ \hspace{0.3em} Radu Florian$^{1}$ \hspace{0.3em} Stacy Patterson$^{2}$ \hspace{0.3em} Ana Milanova$^{2}$ \hspace{0.3em} Wei Sun$^{1}$ \\
$^1$IBM Research\\ $^2$Rensselaer Polytechnic Institute\\
 $^*$ Corresponding author: \texttt{linh@ibm.com}
}

\begin{document}

\maketitle

\begin{abstract}
The strong zero-shot and long-context capabilities of recent Large Language Models (LLMs) have enabled highly effective list-wise reranking systems. Attention-based rerankers leverage transformer attention patterns as retrieval signals, but not all attention heads contribute equally: many introduce noise and redundancy, limiting both reranking quality and efficiency. In this work, we introduce \textit{CoRe heads}, a small subset of retrieval-specialized heads identified through a contrastive scoring objective that rewards attention to relevant documents while penalizing attention to distractors. This contrastive criterion isolates highly discriminative retrieval heads and yields a strong training-free list-wise reranker. Our analysis reveals that retrieval-relevant computation in transformer rerankers is highly sparse and structurally localized: fewer than $1\%$ of attention heads account for most reranking performance across models and datasets, and these heads consistently concentrate in middle transformer layers. This localization enables aggressive layer pruning for efficient reranking: removing the final $50\%$ of transformer layers preserves nearly identical reranking accuracy while substantially reducing inference latency and GPU memory usage. 
\end{abstract}

\section{Introduction}

Information retrieval systems form the backbone of search engines and retrieval-augmented generation (RAG). Modern retrieval pipelines typically use a lightweight retriever such as BM25 \citep{robertson2009probabilistic} or dense retrievers \citep{karpukhin2020dense} to identify candidate documents, followed by a more powerful reranker that refines their ordering. Recent Large Language Models (LLMs) have enabled strong zero-shot reranking capabilities \citep{sachan2022improving, sun2023chatgpt, qin2024large, chenattention}, particularly in list-wise settings where the model jointly reasons over the entire candidate set. 

\begin{wrapfigure}[13]{r}{0.4\textwidth}
\vspace{-0.6cm}
  \begin{center}
    \includegraphics[width=0.4\textwidth]{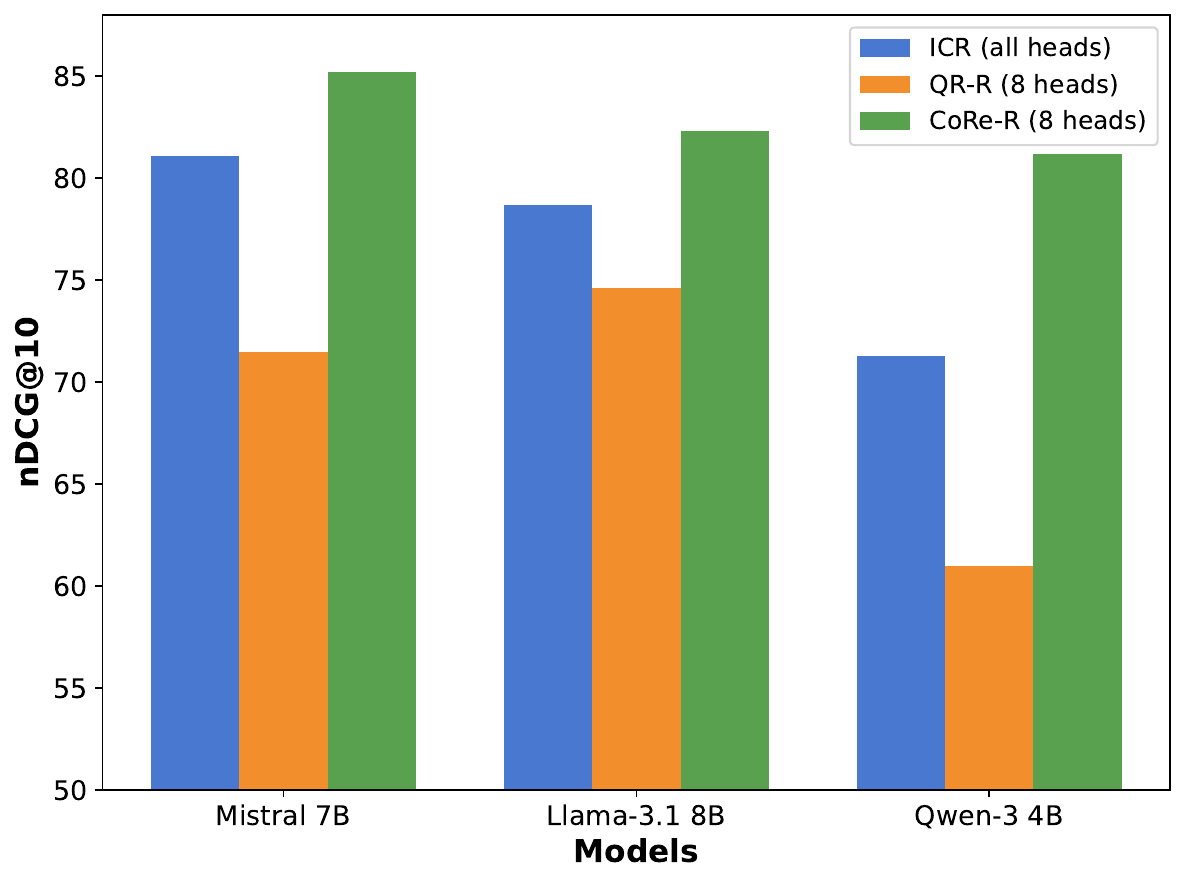}
  \end{center}
  \caption{nDCG@10 on Quora top-40.
  }
  \label{fig.quora}
\end{wrapfigure}

Recent attention-based rerankers exploit transformer attention patterns as reranking signals. The relevance of the documents can be ranked by comparing the cross-attention scores from query tokens to the concatenated document tokens. For example, In-Context Reranking (ICR)~\citep{chenattention} aggregates attention scores across all transformer heads to enable single-pass reranking without autoregressive decoding, while query-focused retrieval  (QR) heads~\citep{zhang2025query} select a subset of heads according to the attention assigned to relevant documents. However, reranking is a fundamentally \emph{contrastive} task: a useful retrieval head should not only attend strongly to relevant documents, but also suppress attention to competing negatives. As a result, heads with high positive attention may still exhibit weak discriminative retrieval behavior. Figure~\ref{fig.quora} illustrates this phenomenon across three different models, where selecting top QR heads can even underperform aggregating attention across all heads as done in ICR.

In this work, we introduce \textit{CoRe heads} (Contrastive Retrieval heads), a small subset of retrieval-specialized attention heads identified using a contrastive scoring objective. Unlike prior retrieval-head selection methods based solely on absolute relevance signals, CoRe explicitly favors heads that discriminate relevant documents from hard negatives. Aggregating signals exclusively from these heads yields a strong training-free reranker while revealing a striking structural property of transformer rerankers: retrieval-specialized heads consistently localize in middle transformer layers, indicating that reranking-relevant computation is highly sparse and structurally concentrated. We further show that this localization can be exploited for aggressive layer pruning, substantially improving inference efficiency with minimal impact on reranking quality. Our contributions are as follows:

\begin{itemize}[leftmargin=*]

\item We propose a training-free contrastive head-selection objective that identifies retrieval-specialized attention heads by jointly rewarding attention to relevant documents and penalizing attention to hard negatives, leading to stronger reranking performance than prior attention-based rerankers.

\item We show that fewer than $1\%$ of attention heads recover and improve reranking quality across diverse retrieval settings. CoRe heads identified using a small subset of Natural Questions generalize across the BEIR benchmark \citep{thakur2beir}, multi-hop retrieval tasks, cross-domain datasets, multilingual retrieval, and long-context settings.

\item We provide the first systematic study of layer pruning in attention-based rerankers. We show that retrieval-specialized heads consistently localize in middle transformer layers, enabling pruning of the final $50\%$ of transformer layers with minimal impact on reranking quality. This reduces GPU memory usage by up to $40\%$ and inference latency by up to $20\%$, demonstrating that reranking-relevant computation is substantially more localized than previously assumed.

\end{itemize}

\section{Related work}

\textbf{Zero-shot reranking.} Zero-shot rerankers with LLMs can be grouped into three main approaches: point-wise, pair-wise, and list-wise. Point-wise methods \citep{sachan2022improving, liang2022holistic} score each document independently, typically through logits or direct generation. These approaches are computationally efficient but often underperform due to the absence of cross-document comparison. Pair-wise methods \citep{qin2024large} compare two documents at a time and aggregate over all pairs to form a ranking. While improving accuracy, the computational cost grows quadratically with the number of documents. List-wise methods \citep{ma2023zero, chenattention} such as RankGPT \citep{sun2023chatgpt} consider all documents jointly to produce a single ranked list. Although list-wise methods require long-context modeling, advances in LLMs with extended context windows \citep{chen2023extending, jin2024llm, fu2024data} have made this increasingly practical. Consequently, list-wise approaches now achieve both scalability and strong effectiveness. Our work follows this trajectory, focusing on attention-based list-wise reranking \citep{chenattention}, which offers more efficiency and stability compared to generation-based methods \citep{ma2023zero, sun2023chatgpt}.

\textbf{Role of attention heads.} A growing body of work in mechanistic interpretability examines how the activation of attention heads shapes model behavior. Previous works \citep{michel2019sixteen, voita2019analyzing} show that only a small fraction of heads are necessary for translation tasks. Olsson et al.~\citep{olsson2022context} identify induction heads that capture repeated input patterns, and later extended the work to heads supporting in-context learning \citep{yinattention, ren2024identifying}. Other studies reveal specialized heads responsible for knowledge conflict \citep{shi2024ircan, jin2024cutting} and context distraction \citep{zhu2025focus}. More recently, Wu et al.~\citep{wu2025retrieval} identify ``Needle In A Haystack'' {(NIAH) heads} associated with copy-and-retrieval behavior in long-context  reasoning tasks, while Zhang et al.~\citep{zhang2025query} propose QR heads that capture query–context interactions beyond direct copying. Collectively, these studies suggest that different subsets of heads specialize in distinct computational roles, and isolating the ones relevant to retrieval remains an open challenge. In this work, we build on the notion of retrieval heads \citep{wu2025retrieval, zhang2025query} as the mechanisms responsible for retrieving relevant information from long-context inputs -- directly aligning with the reranking objective.

\section{Background}
\label{background}
\textbf{Attention-based reranker.}
Given a query $q$ and $k$ candidate documents $D = \{ d_1, d_2, \dots, d_k \}$ returned by a base retriever, the goal of the reranker is to reorder $D$ so that the documents are sorted in the order of relevance to $q$ . To achieve this, Chen et al.~\citep{chenattention} propose In Context Reranking (ICR), an attention-based approach that leverages the attention signal from all attention heads of a LLM to compute fine-grained relevance scores.

Utilizing the long-context capability of LLMs, ICR constructs an input sequence consisting of instruction prompt and $k$ documents followed by query $q$. Let $a_{j,t}^h$ be the post-softmax attention score from the $t$-th token in query $q$ to the $j$-th token in document $d_i$ by an attention head $h$. For each token $j$ in document $d_i$, the token-level relevance score with respect to the query $q$ within a head $h$ is
\begin{align} \label{equation.token.score}
    s^h (d_{i,j}) = \frac{1}{|\mathcal{T}_q|} \sum_{t\in \mathcal{T}_q} a_{j,t}^h,
\end{align}
where $\mathcal{T}_q$ denotes the set of the query tokens.
The document-level relevance score of document $d_i$ within attention head $h$ is aggregated as
\begin{align} \label{equation.document.score}
    s^h(d_i) = \sum_{t\in \mathcal{T}_{d_i}} s^h(d_{i,j})
\end{align}
\noindent where $\mathcal{T}_{d_i}$ denotes the set of the document $d_i$'s tokens.
Aggregating over all heads in the head set $H$, the ICR document-level relevant scores is
\begin{align} \label{equation.icr}
    rel_{ICR}(d_i) = \sum_{h\in H} s^h(d_i).
\end{align}

To reduce bias relative to document length and low-information tokens such as punctuations, ICR applies \textit{contextual calibration} \citep{zhao2021calibrate} that effectively cancels out the redundant or biased attention signals. The calibration process involves two forward passes, one with the actual query and one with a content-free query (e.g. "N/A"), and an outlier filtering process using standard deviation thresholding. When computing document relevance scores, the calibration subtracts the attention score of the content-free query from the attention score of the actual query. Contextual calibration substantially improves the performance of attention-based rerankers while adding little computational overhead by reusing the KV cache.

\textbf{Reranker with QR heads.}
As noted above, ICR computes relevance scores by aggregating attention signals across all attention heads. While comprehensive, such full-head aggregation often introduces redundancy and noise, ultimately degrading reranking performance. Recent work on QR heads \citep{zhang2025query} shows that leveraging only a targeted subset of heads $H_{QR}$ can improve retrieval effectiveness, suggesting that many attention heads are unnecessary for this task. Specifically, the QR document relevance score is computed as
\begin{align} \label{equation.qr}
    rel_{QR}(d_i) = \sum_{h\in H_{QR}} s^h(d_i).
\end{align}
The QR heads $H_{QR}$ is determined via a head detection process on a set of query-documents. For each sample involving a query and $k$ documents, QR rank a head $h$ based on its \textit{retrieval score}
\begin{align}
    S_{QR}(h) = s^h(d^+)
\end{align}
using the same notion of document-level score in Equation~\ref{equation.document.score} for the positive document $d^+$.

The QR scoring metric essentially picks heads with the highest attention to the positive document. However, this metric does not account for the relative ranking within each head's distribution as the document-level scores is not normalized. Because this criterion ignores how strongly a head contrasts the positive document against competing negatives, it fails to capture whether the head meaningfully separates signal from noise. This omission can lead to the selection of misleading heads.

\section{Contrastive retrieval heads}\label{section.core}

Addressing the limitations of prior attention-based reranking approaches, we propose a training-free contrastive head detection framework that identifies a small subset of \emph{Contrastive Retrieval heads} (CoRe heads) specialized for document reranking.

\textbf{Contrastive scoring metric.}
Let $s^h(d_i)$ denote the document-level score of document $d_i$ under attention head $h$ (Equation~\ref{equation.document.score}).
We define the contrastive retrieval score of head $h$ as
\begin{align}\label{s_core.formula}
    S_{CoRe}(h)
    =
    \frac{\exp (s^h(d^+) / t)}
    {\exp (s^h(d^+) / t) + \sum_{i\in D^-} \exp (s^h(d^-_i) / t)},
\end{align}
where $d^+$ is the positive document, $D^-$ is the set of negatives, and $t$ is a temperature hyperparameter.

Intuitively, $S_{CoRe}(h)$ measures the extent to which head $h$ concentrates probability mass on the positive document relative to hard negatives. Heads with larger $S_{CoRe}$ therefore exhibit stronger discriminative ability and are more suitable for reranking. Equivalently, ranking heads by $S_{CoRe}(h)$ is identical to ranking them by the corresponding head-level negative log-likelihood, connecting the proposed criterion to contrastive objectives such as InfoNCE~\citep{oord2018representation}.

\textbf{Identifying CoRe heads.}
We compute the $S_{CoRe}$ of each attention head averaged over a small set of data samples, and select the top-scoring few attention heads within each model as CoRe heads. By aggregating attention signals from these CoRe heads, the reranker achieves improved performance compared to aggregating over all heads.

\textbf{Hard negative data.} 
We compute the retrieval score $S_{CoRe}$ for all attention heads using a subset of $1000$ samples from the Natural Questions (NQ) training set \citep{kwiatkowski2019natural}. Each sample contains a query, a positive answer passage (gold document), and $49$ hard negatives mined with the \emph{ibm-granite/granite-embedding-30m-english} model \citep{awasthy2025graniteembeddingmodels}. Hard negatives are constructed by sampling from the top 100 candidates returned by the retriever system. To further reduce false negatives, we discard any passage whose similarity to the query exceeds that of the gold passage \citep{moreira2025nvretrieverimprovingtextembedding}. This procedure ensures that $S_{CoRe}$ is computed under challenging retrieval conditions, ensuring that the detected heads are more discriminative and robust for reranking tasks.

\textbf{Head detection process.}\label{section.head.detection} For each sample, we construct a LLM prompt consisting of the $50$ documents, followed by the instruction and query (see Appendix~\ref{app.prompt} for details). The positive document is inserted among the first five different positions. We compute the $S_{CoRe}$ of each attention head on each sample, and compute the final retrieval head score by averaging across all samples. Less than top $1\%$ of all heads with the highest averaged score are selected as CoRe heads ($8$ heads in our experiment). 
This setup yields stable average retrieval scores, with highly overlapping top-ranked CoRe heads across repeated evaluations for each model.
Moreover, CoRe heads identified using a single dataset NQ generalize effectively across multiple datasets, enhancing reranking performance as shown later in Section~\ref{section.exp.result}. The detection process is light-weight, training-free and relatively fast, requiring less than one hour on an H100 (96GB) GPU for the models considered in this work.

\textbf{Reranking with CoRe heads.} Once the set of CoRe heads are identified for a given language model, zero-shot reranking is performed by aggregating attention signals exclusively from this subset. Specifically, the CoRe document relevance score is
\begin{align}
    rel_{CoRe}(d_i) = \sum_{h\in H_{CoRe}} s^h(d_i),
\end{align}
where $H_{CoRe}$ denotes the set of detected CoRe heads.

\textbf{Stability of post-hoc sparse head detection.} Because CoRe head selection is based on an additive ranking objective over individual head scores, the optimal head subset is obtained by selecting the top-$m$ heads according to their population-level CoRe scores. The following result characterizes when finite-sample estimation recovers the same sparse retrieval head set with high probability.

\begin{proposition}
\label{thm:gap} \textbf{(Gap-dependent stability of CoRe head detection)}
Let $\mathcal{H}$ be a finite set of attention heads. For each head $h\in\mathcal{H}$, define the population and empirical CoRe scores
\[
\mu_h
=
\mathbb{E}_{Z\sim\mathcal{D}}
[S_{\mathrm{CoRe}}(h;Z)],
\qquad
\hat{\mu}_h
=
\frac{1}{N}
\sum_{i=1}^N
S_{\mathrm{CoRe}}(h;Z_i),
\]
where $Z_1,\dots,Z_N\sim\mathcal{D}$ are i.i.d.\ reranking instances and
$S_{\mathrm{CoRe}}(h;Z)\in[0,1]$.

Let $A^*$ and $\hat{A}$ denote the top-$m$ heads under $\mu_h$ and $\hat{\mu}_h$, respectively. Suppose the population scores satisfy the separation condition
\[
\Delta
=
\min_{h\in A^*,\,h'\notin A^*}
(\mu_h-\mu_{h'})
>0.
\]
Then with probability at least
\[
1
-
2|\mathcal{H}|
\exp\left(
-\frac{N\Delta^2}{2}
\right),
\]
the empirical top-$m$ head set exactly recovers the population top-$m$ head set: i.e., 
$\hat{A}=A^*$.
\end{proposition}

Proposition~\ref{thm:gap} shows that post-hoc CoRe head detection is stable under the additive top-$m$ ranking objective whenever retrieval-specialized heads are sufficiently separated from the remaining heads under the population-level CoRe objective. Although the result applies broadly to additive attention-based head ranking objectives, it establishes an important structural property of sparse head detection: retrieval-specialized heads can be stably recovered from finite samples whenever sufficiently separated from the remaining heads. Empirically, we observe substantial score separation between selected and non-selected CoRe heads, helping explain the reproducibility of detected head sets across datasets and hard-negative configurations.

\textbf{Comparison to QR-based head selection.}
The previous result characterizes when sparse retrieval-specialized heads can be stably identified from finite samples. We now consider a complementary question: why does the contrastive CoRe objective produce more discriminative retrieval heads than positive-only QR scoring? Intuitively, QR does not penalize heads that allocate substantial attention to irrelevant documents. The following proposition formalizes this limitation by showing that larger attention to the positive document alone does not guarantee stronger discriminative retrieval behavior.

\begin{proposition}\label{prop:qr_suboptimal}
\textbf{(Positive-only attention does not guarantee discriminative retrieval behavior)}
Fix temperature $t>0$ and any candidate set containing one positive document $d^+$ and at least one negative document. There exist two attention heads $a$ and $b$ such that
\[
s^a(d^+) > s^b(d^+), \quad\text{but}\quad
S_{CoRe}(a) < S_{CoRe}(b).
\]
\end{proposition}

This result highlights that retrieval-specialized heads should be evaluated not only by the attention assigned to relevant documents, but also by their ability to distinguish relevant documents from distractors. Consistent with the theoretical analysis, CoRe produces larger empirical margins between consecutively ranked head scores than QR across most head ranks (Appendix~\ref{app.head.score.gap}), indicating improved discrimination between informative and non-informative heads.

\section{Experiments}
\label{section.exp}

In this section, we evaluate the effectiveness of CoRe heads in reranking task across a diverse set of datasets and open-weight decoder models.
Our code is available at \url{https://github.com/linhhtran/CoRe-Reranking}.

\subsection{Setup}\label{section.exp.setup}

\textbf{Baselines.} We compare CoRe-R against existing zero-shot attention-based rerankers, including ICR \citep{chenattention} and QR-R \citep{zhang2025query}. For QR head detection, we use the same detection dataset described in Section~\ref{section.head.detection} to ensure a fair comparison. In addition, we include comparisons with NIAH heads in selected localization (Appendix~\ref{app.head.location}).

\textbf{Datasets.} We evaluate our approach on the BEIR benchmark \citep{thakur2beir}. We also evaluate MLDR \citep{chen2024bge}, MuSiQue \citep{trivedi2022musique} and BRIGHT \citep{hongjinbright} as out-of-domain datasets. We evaluate the reranking tasks using a diverse set of base retrievers (BM25, dense or ColBERT) which are detailed in Appendix~\ref{app.config}. For each dataset, we rerank top-40 documents starting from the original retrieved order (see Appendix~\ref{app.doc.order} for experiments on different document orders).

\textbf{Language models.} Since attention-based reranking requires access to all attention scores of all layers and heads, we focus on open-source LLMs. Specifically, we evaluate \textit{mistralai/Mistral-7B-Instruct-v0.2} \citep{jiang2023mistral7b}, \textit{meta-llama/Llama-3.1-8B-Instruct} \citep{dubey2024llama}, \textit{Qwen/Qwen3-4B-Instruct-2507} \citep{qwen3embedding} and \textit{ibm-granite/granite-3.2-8b-instruct} \citep{granitehf}. To ensure consistency, we adopt a uniform prompt structure for all datasets (see Appendix~\ref{app.prompt}) and apply the same contextual calibration to all rerankers.

\textbf{Hyperparameters.} The contrastive temperature $t$ is tuned on a separate subset of the NQ train dataset. We select $t=0.001$ for Mistral and Granite, and $t=0.1$ for Llama and Qwen. These values are fixed for CoRe-R head detection prior to running the zero-shot reranking experiments. Both QR-R and CoRe-R use the top $8$ retrieval heads, and report performance using the nDCG@10 scores.

\setlength{\tabcolsep}{4pt}
\begin{table}[t]
\caption{nDCG@10 on the BEIR benchmark top-$40$.}
\label{table.beir}
\begin{center}
\scriptsize
\begin{tabular}{ | c | c | c c c | c c c | c c c | c c c | }
\hline
Task & Dense & \multicolumn{3}{c|}{Mistral 7B} & \multicolumn{3}{c|}{Llama-3.1 8B} & \multicolumn{3}{c|}{Qwen-3 4B} & \multicolumn{3}{c|}{Granite-3.2 8B} \\\cline{3-14}
& baseline & ICR & QR-R & CoRe-R & ICR & QR-R & CoRe-R & ICR & QR-R & CoRe-R & ICR & QR-R & CoRe-R \\
\hline\hline
TREC-COVID & 63.1 & 72.1 & 73.1 & \textbf{73.8} & \textbf{76.8} & 75.7 & 75.7 & 74.2 & 75.5 & \textbf{76.4} & 70.4 & 70.3 & \textbf{71.8} \\
NFCorpus & 33.7 & 33.1 & 35.0 & \textbf{35.2} & 35.0 & 36.4 & \textbf{36.7} & 35.2 & 32.4 & \textbf{36.7} & 31.8 & 32.5 & \textbf{33.9} \\
DBPedia & 36.0 & 36.0 & 36.8 & \textbf{37.5} & 38.7 & 39.0 & \textbf{39.4} & 38.4 & 36.2 & \textbf{39.3} & 35.6 & 35.9 & \textbf{36.8} \\
SciFact & 71.3 & 71.3 & 71.9 & \textbf{74.7} & 74.7 & \textbf{75.9} & 75.1 & 72.1 & 73.4 & \textbf{76.1} & 74.6 & 74.7 & \textbf{75.2} \\
SciDocs & 22.5 & 16.6 & 17.3 & \textbf{19.5} & 18.5 & 19.7 & \textbf{20.2} & 20.8 & 20.3 & \textbf{21.6} & 18.4 & \textbf{20.3} & 20.0 \\
FiQA & 36.9 & 35.8 & 40.3 & \textbf{41.4} & 41.3 & 44.0 & \textbf{44.1} & 40.2 & 41.4 & \textbf{44.0} & 39.7 & 41.6 & \textbf{42.4} \\
NQ & 51.6 & 55.4 & 56.5 & \textbf{58.1} & 60.8 & 63.0 & \textbf{63.2} & 61.5 & 61.8 & \textbf{64.1} & 57.4 & 57.9 & \textbf{60.7} \\
FEVER & 85.5 & 87.5 & 87.1 & \textbf{88.7} & \textbf{89.2} & 87.4 & 88.4 & 87.1 & 87.2 & \textbf{88.5} & \textbf{88.0} & 87.5 & 86.2 \\
Climate-FEVER & 30.3 & 21.6 & 21.1 & \textbf{23.3} & 22.5 & \textbf{22.9} & 22.7 & 22.5 & 21.3 & \textbf{24.5} & 20.5 & \textbf{21.2} & 21.0 \\
HotpotQA & 62.9 & 71.7 & 71.6 & \textbf{73.4} & 73.7 & 73.6 & \textbf{73.8} & 72.3 & 73.5 & \textbf{74.0} & \textbf{73.6} & 72.8 & 72.6 \\
Touche & 24.0 & 23.3 & 26.4 & \textbf{26.7} & 26.1 & 26.3 & \textbf{26.4} & \textbf{27.2} & 27.0 & \textbf{27.2} & 20.5 & 24.6 & \textbf{24.8} \\
MSMARCO & 30.7 & 30.1 & \textbf{32.1} & 31.7 & 32.9 & 33.9 & \textbf{34.9} & 33.2 & 33.5 & \textbf{34.4} & 29.6 & 31.2 & \textbf{32.5} \\
Quora & 86.7 & 81.1 & 71.5 & \textbf{85.2} & 78.7 & 74.6 & \textbf{82.3} & 71.3 & 61.0 & \textbf{81.2} & \textbf{83.1} & 79.8 & 80.1 \\
ArguAna & 56.4 & 45.0 & 51.1 & \textbf{52.7} & 42.5 & 54.2 & \textbf{55.0} & 39.7 & 41.4 & \textbf{52.9} & 52.3 & 56.1 & \textbf{57.3} \\
CQADupstack & 44.3 & 38.7 & 40.8 & \textbf{41.6} & 41.7 & 43.2 & \textbf{43.9} & 42.6 & 42.5 & \textbf{44.8} & 41.4 & 42.3 & \textbf{43.4} \\
\hline
Average & 49.1 & 47.9 & 48.9 & \textbf{50.9} & 50.2 & 51.3 & \textbf{52.1} & 49.2 & 48.6 & \textbf{52.4} & 49.1 & 50.0 & \textbf{51.0} \\
\hline
\end{tabular}
\end{center}
\end{table}

\subsection{Reranking results}
\label{section.exp.result}

\begin{figure*}[b]
\begin{minipage}[t]{0.32\textwidth}
  \includegraphics[width=\linewidth]{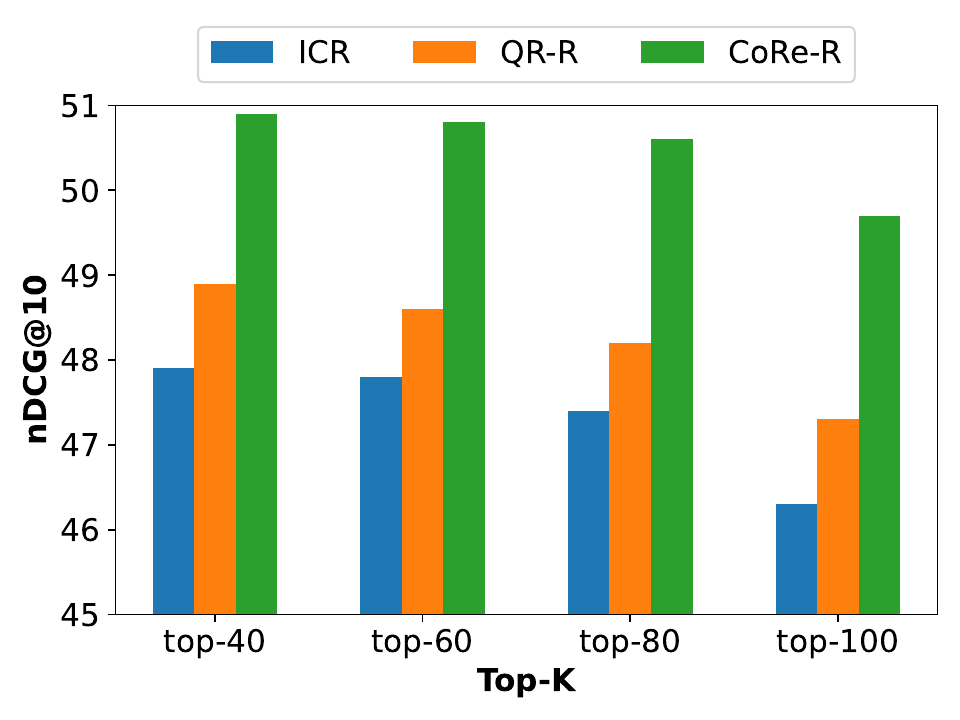}
  \caption{Average nDCG@10 on BEIR with various top-$k$\\ using Mistral 7B.}\label{fig.long.context.mistral}
\end{minipage}%
\hspace{2pt}
\begin{minipage}[t]{0.32\textwidth}
  \includegraphics[width=\linewidth]{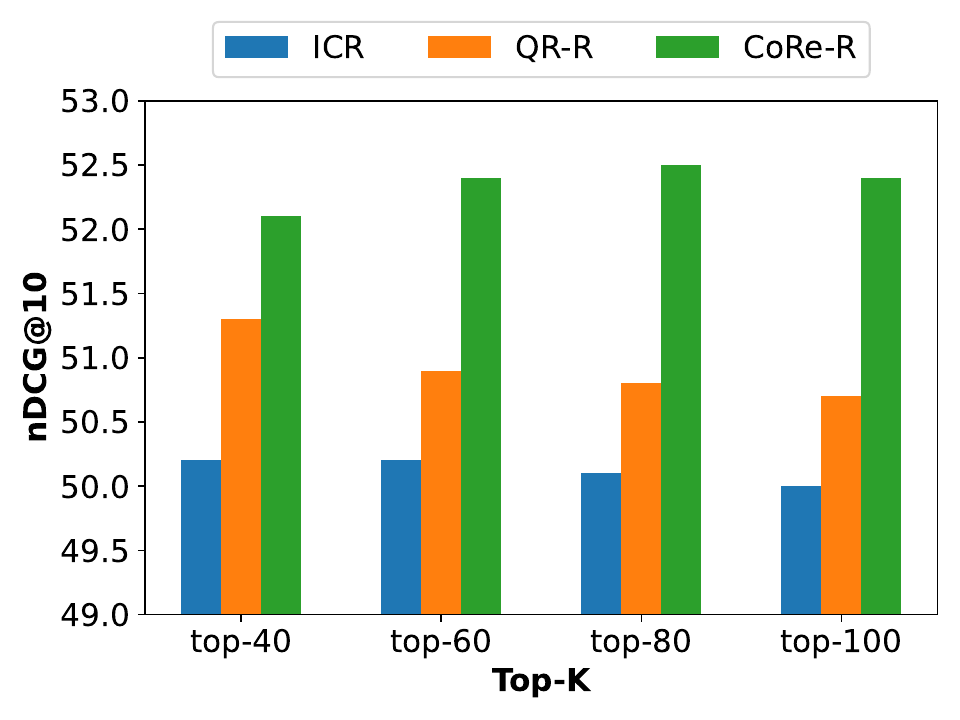}
  \caption{Average nDCG@10 on BEIR with various top-$k$\\ using Llama-3.1 8B.}\label{fig.long.context.llama}
\end{minipage}%
\hspace{2pt}
\begin{minipage}[t]{0.32\textwidth}
  \includegraphics[width=\linewidth]{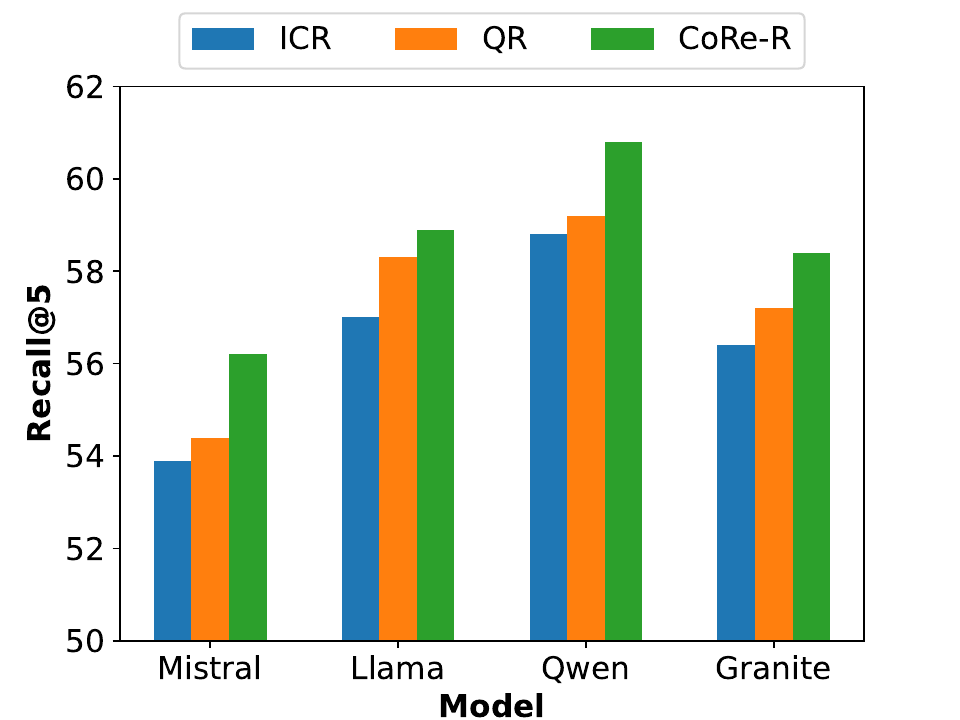}
  \caption{Recall@5 on multi-hop dataset MuSiQue top-$40$.}\label{fig.musique}
\end{minipage}%
\end{figure*}

\textbf{Reranking performance on BEIR.} 
Table~\ref{table.beir} reports reranking results on the BEIR benchmark. Across all four models, CoRe-R achieves the strongest performance, consistently outperforming both ICR and QR-R. Aggregating attention from all heads in ICR yields the weakest results, while CoRe-R provides clear and uniform improvements over both the retriever and ICR baselines. Relative to QR-R, CoRe-R shows the largest averaged gains on Mistral ($+2.0$) and Qwen ($+3.8$) and delivers consistent improvements on both Llama and Granite. We report additional results using BM25 retriever in Appendix~\ref{app.beir.bm25}.

We observe that CoRe-R shows remarkable improvement on Quora compared to all baselines in Table~\ref{table.beir}. Quora is a duplicate-question retrieval task with a high proportion of hard negatives.
In such cases, the contrastive metric shines as it explicitly rewards heads that emphasize the positive document while suppressing near-duplicate.
As a result, CoRe-R performs extremely well, while QR-R shows a drastic decline across all models. This aligns well with our theoretical analysis in Proposition~\ref{prop:qr_suboptimal}, demonstrating that CoRe heads is more optimal.

\textbf{Long context scalability.}
Figure~\ref{fig.long.context.mistral} and Figure~\ref{fig.long.context.llama} report the average accuracy on BEIR benchmark under varying context lengths (see Appendix~\ref{app.long.context} for detailed results). 
Across all context settings, CoRe-R consistently outperforms both ICR and QR-R.
For Mistral, all methods experience noticeable degradation as the context length increases. This trend aligns with prior observations: model pre-trained with limited context length (Mistral at 16k) is not robust under long-context settings, e.g., attempts to extend its context window through long-context fine-tuning have been shown to degrade model quality~\citep{scale2024longcontext} (see Appendix~\ref{app.long.context} for another example on Phi model). For Llama, both baselines ICR and QR-R show mild degradation at longer context lengths. In contrast, CoRe-R continues to improve as context grows, further widening the performance gap. 
This behavior highlights the strong scalability of CoRe heads to longer retrieval contexts.

\setlength{\tabcolsep}{4pt}
\begin{table}[t]
\caption{nDCG@10 on BRIGHT top-$40$.}
\label{table.bright}
\begin{center}
\scriptsize
\begin{tabular}{ | c | c | c c c | c c c | c c c | }
\hline
Task & BM25 & \multicolumn{3}{c|}{Mistral 7B} & \multicolumn{3}{c|}{Llama-3.1 8B} & \multicolumn{3}{c|}{Qwen-3 4B} \\\cline{3-11}
& baseline & ICR & QR-R & CoRe-R & ICR & QR-R & CoRe-R & ICR & QR-R & CoRe-R \\
\hline\hline
Bio.    & 19.2 &    33.3 & 32.2 & \textbf{40.0} &     39.1 & 38.7 & \textbf{40.7}   & 43.8 & \textbf{44.8} & \textbf{44.8} \\
Earth.  & 27.1 &    38.9 & 39.4 & \textbf{43.7} &     38.3 & 40.2 & \textbf{42.1}   & 46.0 & 46.5 & \textbf{47.2} \\
Econ.   & 14.9 &    20.0 & 21.8 & \textbf{22.4} &     21.1 & 25.1 & \textbf{26.0}   & 16.9 & 17.0 & \textbf{26.6} \\
Psy.    & 12.5 &    32.8 & 33.2 & \textbf{33.5} &     33.2 & 36.7 & \textbf{38.3}   & 33.8 & 38.8 & \textbf{40.6} \\
Rob.    & 13.5 &    18.2 & 19.3 & \textbf{20.6} &     17.3 & 21.1 & \textbf{22.0}   & 20.3 & 22.6 & \textbf{25.0} \\
Stack.  & 16.5 &    21.2 & 21.8 & \textbf{22.7} &     19.6 & 23.0 & \textbf{23.6}   & 21.4 & 24.3 & \textbf{26.4} \\
Sus.    & 15.2 &    21.2 & 23.9 & \textbf{25.4} &     22.1 & 24.9 & \textbf{26.5}   & 24.5 & 25.4 & \textbf{26.3} \\
Pony    &  7.9 &     4.4 &  6.2 &  \textbf{7.7} &      5.5 &  8.4 & \textbf{10.4}   & 11.4 & 11.3 & \textbf{12.2} \\
Leet.   & 24.4 &    24.6 & 26.5 & \textbf{27.3} &     23.5 & 26.9 & \textbf{28.3}   & 23.5 & 23.7 & \textbf{29.0} \\
AoPS    &  6.0 &     7.6 &  \textbf{8.4} &  8.0 &      7.5 &  \textbf{8.4} &  8.2   &  7.2 &  8.3 &  \textbf{8.6} \\
TheoQ   & 13.0 &    18.9 & 20.3 & \textbf{21.8} &     18.7 & 22.4 & \textbf{24.1}   & 20.5 & 23.9 & \textbf{24.6} \\
TheoT   &  6.9 &     9.8 & 11.0 & \textbf{12.1} &     11.0 & 14.6 & \textbf{16.1}   & 15.1 & 17.6 & \textbf{20.5} \\
\hline
Average & 14.8 &    20.9 & 22.0 & \textbf{23.8} &     21.4 & 24.2 & \textbf{25.5}   & 23.7 & 25.3 & \textbf{27.7} \\
\hline
\end{tabular}
\end{center}
\end{table}

\setlength{\tabcolsep}{3pt}
\begin{table}
\parbox{.48\linewidth}{
\caption{nDCG@10 on MLDR top-$40$.}
\label{table.mldr}
\centering
\scriptsize
\begin{tabular}{ | c | c | c c c | c c c | }
\hline
Lang. & Dense & \multicolumn{3}{c|}{Mistral 7B} & \multicolumn{3}{c|}{Llama-3.1 8B} \\\cline{3-8}
& baseline & ICR & QR-R & CoRe-R & ICR & QR-R & CoRe-R \\
\hline\hline
Ger. & 19.9        & 23.7 & 27.2 & \textbf{28.2}         & 24.9 & 27.8 & \textbf{28.6} \\
Eng. & 29.5       & 23.7 & 24.6 & \textbf{28.1}         & 29.1 & 29.7 & \textbf{30.2} \\
Spa. & 43.3       & 39.1 & 43.3 & \textbf{45.8}         & 43.3 & 46.1 & \textbf{48.0} \\
Fre. & 49.1        & 47.3 & 46.9 & \textbf{53.0}         & 50.1 & 52.3 & \textbf{53.6} \\
Ita. & 41.9       & 36.0 & 35.9 & \textbf{42.5}         & 41.6 & 43.1 & \textbf{43.7} \\
Por. & 52.1    & 46.1 & 49.1 & \textbf{54.7}         & 52.3 & 55.2 & \textbf{57.1} \\
\hline
Avg. & 39.3       & 36.0 & 37.8 & \textbf{42.1}         & 40.2 & 42.4 & \textbf{43.5} \\
\hline
\end{tabular}
}
\hfill
\parbox{.48\linewidth}{
\caption{Win rate and nDCG@10 on extreme cases in BEIR.}
\label{table.length.bias}
\scriptsize
\centering
\begin{tabular}{| c | c | c c c c |}
    \hline
    \multicolumn{2}{|c|}{Reranker} & Win rate & Win rate & nDCG@10 & nDCG@10 \\
    \multicolumn{2}{|c|}{} & (w/o calib) & (w/ calib) & (w/o calib) & (w/ calib) \\
    \hline\hline
    \parbox[t]{2mm}{\multirow{3}{*}{\rotatebox[origin=c]{90}{Mistral}}} & ICR & 27.27 & 57.39 & 10.03 & 21.77 \\
    & QR-R & 50.00 & 68.02 & 22.86 & 25.52 \\
    & CoRe-R & \textbf{56.02} & \textbf{71.91} & \textbf{23.29} & \textbf{27.68} \\
    \hline
    \parbox[t]{2mm}{\multirow{3}{*}{\rotatebox[origin=c]{90}{Llama}}} & ICR & 31.25 & 59.09 & 10.81 & 25.00 \\
    & QR-R & 57.39 & 67.89 & 25.67 & 28.75 \\
    & CoRe-R & \textbf{57.89} & \textbf{69.32} & \textbf{26.92} & \textbf{29.02} \\
    \hline
\end{tabular}
}
\end{table}

\textbf{Out-of-domain generalization.} Table~\ref{table.bright} shows accuracy on BRIGHT and Table~\ref{table.mldr} reports the nDCG@10 on MLDR.
With the same set of CoRe heads detected using a subset of the NQ dataset, CoRe-R shows major improvement in the reranking accuracy over all baselines.
Figure~\ref{fig.musique} depicts the Recall@5 score on the multi-hop dataset MuSiQue, and CoRe-R continues to deliver the highest accuracy across all models (see Appendix~\ref{app.multihop} for more multi-hop results).
Overall, CoRe-R demonstrates consistent gains on cross-lingual, cross-domain and multi-hop tasks, underscoring the strong generalization ability of CoRe heads.

\begin{wraptable}[9]{r}{0.4\textwidth}
\vspace{-0.5cm}
\scriptsize
\centering
\caption{nDCG@10 on BEIR top-40.}
\label{table:llama:70b}
\begin{tabular}{ | c | c | c c c | }
\hline
Task & Dense & \multicolumn{3}{c|}{Llama-3.3 70B} \\\cline{3-5}
& baseline & ICR & QR-R & CoRe-R \\
\hline\hline
TREC-COVID  & 63.1      & 74.2 & 75.0 & \textbf{75.9} \\
NFCorpus    & 33.7      & 30.5 & 34.7 & \textbf{35.4} \\
DBPedia     & 36.0      & 36.3 & 37.4 & \textbf{39.8} \\
SciFact     & 71.3      & 68.0 & 73.4 & \textbf{75.1} \\
Scidocs     & 22.5      & 15.4 & 17.0 & \textbf{20.3} \\
FiQA        & 36.9      & 36.8 & 42.5 & \textbf{44.6} \\
\hline
Average     & 43.9      & 43.5 & 46.6 & \textbf{48.5} \\
\hline
\end{tabular}
\end{wraptable}

\textbf{Larger model scale.}
To examine how well the selected CoRe heads scale with much larger decoder-only models, we evaluate CoRe-R using \textit{meta-llama/Llama-3.3-70B-Instruct} on a subset of BEIR and report the reranking results in Table~\ref{table:llama:70b}. Even though a large-scale model exhibits stronger prompt understanding capabilities, in-context baseline ICR underperforms the dense retriever baseline in several tasks. Meanwhile, CoRe-R scales strongly to this substantially larger model, outperforming both baselines across all evaluated tasks.

\textbf{Robustness against document length bias.}
As discussed in Section~\ref{background}, attention-based rerankers can exhibit document length bias, which may be mitigated through contextual calibration. To evaluate both the severity of this bias and the effectiveness of calibration, we construct a challenging subset of over $4000$ BEIR examples satisfying an extreme length imbalance condition: the ratio between the gold document length and the longest irrelevant document is at most $0.2$ (details in Appendix~\ref{app.length.bias}).

Table~\ref{table.length.bias} reports the win rate at which the gold document receives a higher score than the longest irrelevant document, together with the overall nDCG@10 performance for different attention-based rerankers. Without calibration, ICR suffers substantially from length bias, whereas QR-R and CoRe-R exhibit noticeably smaller degradation. We hypothesize that sparse head selection itself acts as an implicit form of calibration by filtering out weak heads that contribute redundant or length-biased signals. When contextual calibration is applied, CoRe-R achieves the highest win rate and delivers the strongest reranking performance overall.

\section{Stability of CoRe head selection}
\label{section.exp.stability}

\textbf{Stability of selected CoRe heads. }
We evaluate the stability of the selected CoRe heads by repeating the head detection using calibration data from substantially different domains: biomedical/scientific retrieval (BioASQ and SciFact), finance (FiQA), and fact-checking (FEVER), in addition to the original NQ calibration set. Among these datasets, BioASQ is out-of-domain from BEIR benchmark.

Across these calibration datasets, the detected top-eight head sets remain highly stable: all alternatives share 6 of 8 heads, and pair-wise overlap with NQ reaches 7 of 8. Using these head sets for CoRe-R, we report the average nDCG@10 on BEIR in Table~\ref{table.head.stability}, and the average BEIR performance varies by at most 0.3 for either model.

Throughout our experiments, we detect the top-8 heads using $N=1000$ unique samples. We now investigate how much the selected set changes with $N$. Table~\ref{table.score.gap} reports the average number of heads shared between two independent draws (overlap, out of 8), how often the two draws differ, and — when they differ — the correlation between the dropped head and its best-matched replacement.

The overlap between two independent calibration draws rises with $N$ and plateaus by 800-1000, so adding more samples beyond that does not change the selection — confirming the scale we used. The residual variation is a single head: two draws differ by one boundary head on average, and that head is a near-equivalent of the one it replaces, with per-question score correlation 0.80. The 8th and 9th heads are also near-tied in score (the gap is $4e-4$ to $6e-3$). So a different calibration draw only exchanges one boundary head for a redundant near-duplicate, which the aggregated reranker is insensitive to.

\setlength{\tabcolsep}{3pt}
\begin{table}
\parbox{.48\linewidth}{
\caption{Average nDCG@10 on BEIR top-$40$ using different detection datasets.}
\label{table.head.stability}
\centering
\scriptsize
\begin{tabular}{ | c | c c c | }
\hline
Detection & \multicolumn{3}{c|}{Model} \\\cline{2-4}
dataset & Mistral 7B & Llama-3.1 8B & Qwen-3 4B \\
\hline\hline
NQ (original)   & 50.9 & 52.1 & 52.4 \\
BioASQ          & 50.6 & 51.9 & 52.3 \\
SciFact         & 50.7 & 51.8 & 52.1 \\
FiQA            & 50.8 & 51.8 & 52.3 \\
FEVER           & 50.9 & 52.0 & 52.2 \\
\hline
\end{tabular}
}
\hfill
\parbox{.48\linewidth}{
\caption{CoRe head overlap by the number of detection samples (Llama-3.1 8B).}
\label{table.score.gap}
\scriptsize
\centering
\begin{tabular}{| c | c c c |}
    \hline
    Number of & Overlap out of & Number of & Score swap \\
    unique samples & top-8 heads & heads differing & correlation \\
    \hline\hline
    100 & 6.5 & 1.7 & 0.79 \\
    200 & 6.8 & 1.4 & 0.80 \\
    400 & 7.1 & 1.1 & 0.80 \\
    600 & 7.3 & 1.0 & 0.80 \\
    800 & 7.3 & 1.0 & 0.80 \\
    1000 & 7.3 & 1.0 & 0.80 \\
    \hline
\end{tabular}
}
\end{table}

\textbf{Robustness of the contrastive metric. }
We investigate the effectiveness and robustness of the InfoNCE-inspired CoRe metrics by comparing it to three simple alternative metrics for each head $h$: ratio ($s^h(d^+) / s^h(d^-)$), difference ($s^h(d^+) - s^h(d^-)$) and pairwise comparison. For each criterion, we selected the top 8 heads and evaluated the resulting reranker on BEIR.

\begin{wraptable}[8]{r}{0.4\textwidth}
\vspace{-0.5cm}
\scriptsize
\centering
\caption{Average nDCG@10 on BEIR top-40 using heads detected with different metrics.}
\label{table:alternative:metrics}
\begin{tabular}{ | c | c c | }
\hline
Metric & Mistral 7B & Qwen-3 4B \\
\hline\hline
InfoNCE (CoRe) & 50.9 & 52.4 \\
Ratio & 50.6 & 18.0 \\
Difference & 48.6 & 49.8 \\
Pairwise & 48.3 & 50.2 \\
\hline
\end{tabular}
\end{wraptable}

The reranking results are reported in Table~\ref{table:alternative:metrics}. The difference and pairwise criteria underperform CoRe on both models, showing that the positive-vs-negative alternatives select weaker heads. The ratio criterion is competitive on Mistral (50.6), but collapses on Qwen (18.0). This collapse is not an anomaly, but a structural weakness of the ratio $s^h(d^+) / s^h(d^-)$ that diverges whenever a head assigns near-zero attention to negatives. Therefore, the criterion can rank a head arbitrarily high on the basis of a vanishing denominator, rather than discriminative separation. The InfoNCE-style objective of CoRe normalizes the positive attention against the full set of competing candidates, which bounds the score and avoids this instability.

\section{Sparse retrieval specialization enables efficient reranking}
\label{section.exp.prune}

Although CoRe heads are identified using only a contrastive reranking objective, we observe consistent structural patterns across models and datasets. In this section, we analyze these retrieval-specialized heads and show how their sparse organization enables efficient reranking.

\begin{figure*}[t]
\begin{minipage}[t]{0.48\textwidth}
  \includegraphics[scale=0.35]{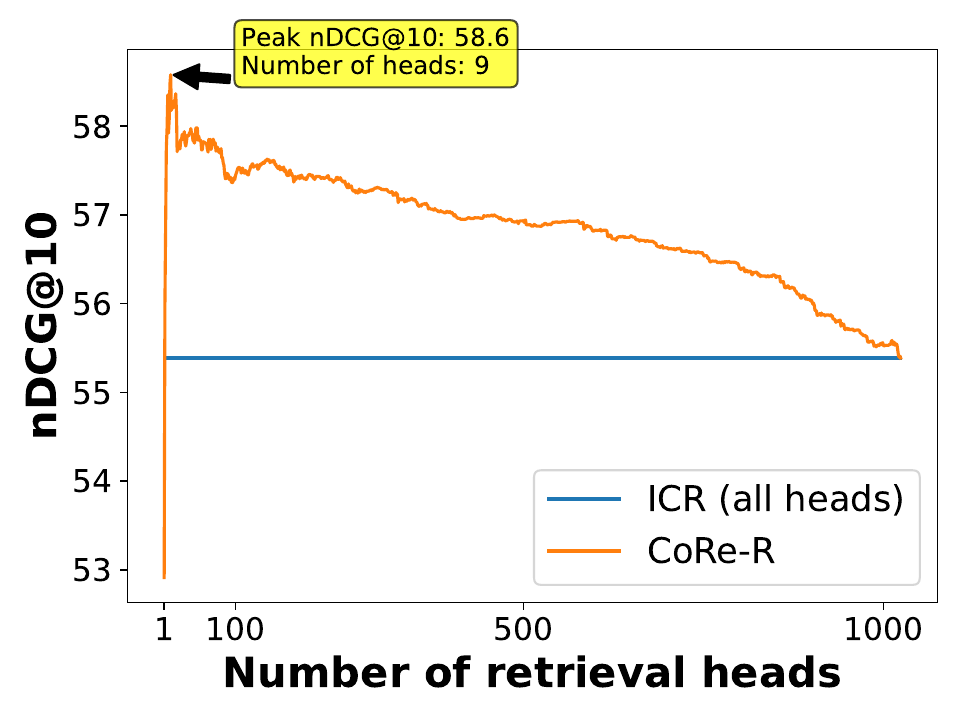}
  \caption{Reranking accuracy of NQ using \\ different top CoRe heads in Mistral.}\label{fig.mistral.nq}
\end{minipage}%
\hspace{3pt}
\begin{minipage}[t]{0.48\textwidth}
  \includegraphics[scale=0.35]{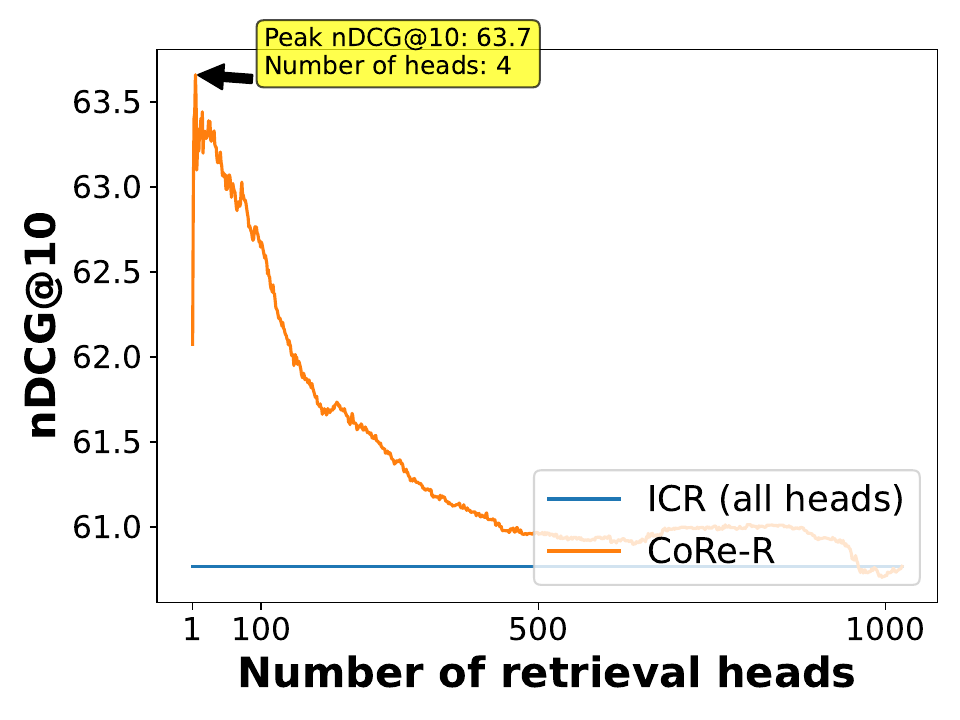}
  \caption{Reranking accuracy of NQ using \\ different top CoRe heads in Llama.}\label{fig.llama.nq}
\end{minipage}%
\end{figure*}

\textbf{Retrieval signal concentrates in a small number of heads. }
Across all models and datasets, we find that optimal reranking performance is typically achieved using fewer than $10$ CoRe heads. Figure~\ref{fig.mistral.nq} and Figure~\ref{fig.llama.nq} show the effect of aggregating different numbers of top-ranked CoRe heads on reranking accuracy for the NQ dataset. In both models, reranking performance peaks with only a handful of retrieval-specialized heads: 9 CoRe heads for Mistral and 4 CoRe heads for Llama.

Notably, adding additional lower-ranked heads often reduces reranking accuracy, suggesting that retrieval-relevant computation is highly concentrated while weaker heads introduce redundant or noisy retrieval signals. This behavior indicates that only a small subset of transformer heads contributes meaningfully to reranking quality. Similar trends are observed on additional datasets (FiQA and NFCorpus), as reported in Appendix~\ref{app.optimal.core.head}.

\begin{wrapfigure}[11]{r}{0.5\textwidth}
\vspace{-0.7cm}
    \centering
    \hspace{-0.5cm}
    \begin{minipage}{0.49\linewidth}
        \includegraphics[scale=0.17,trim={0 0.5cm 0 0.5cm},clip]{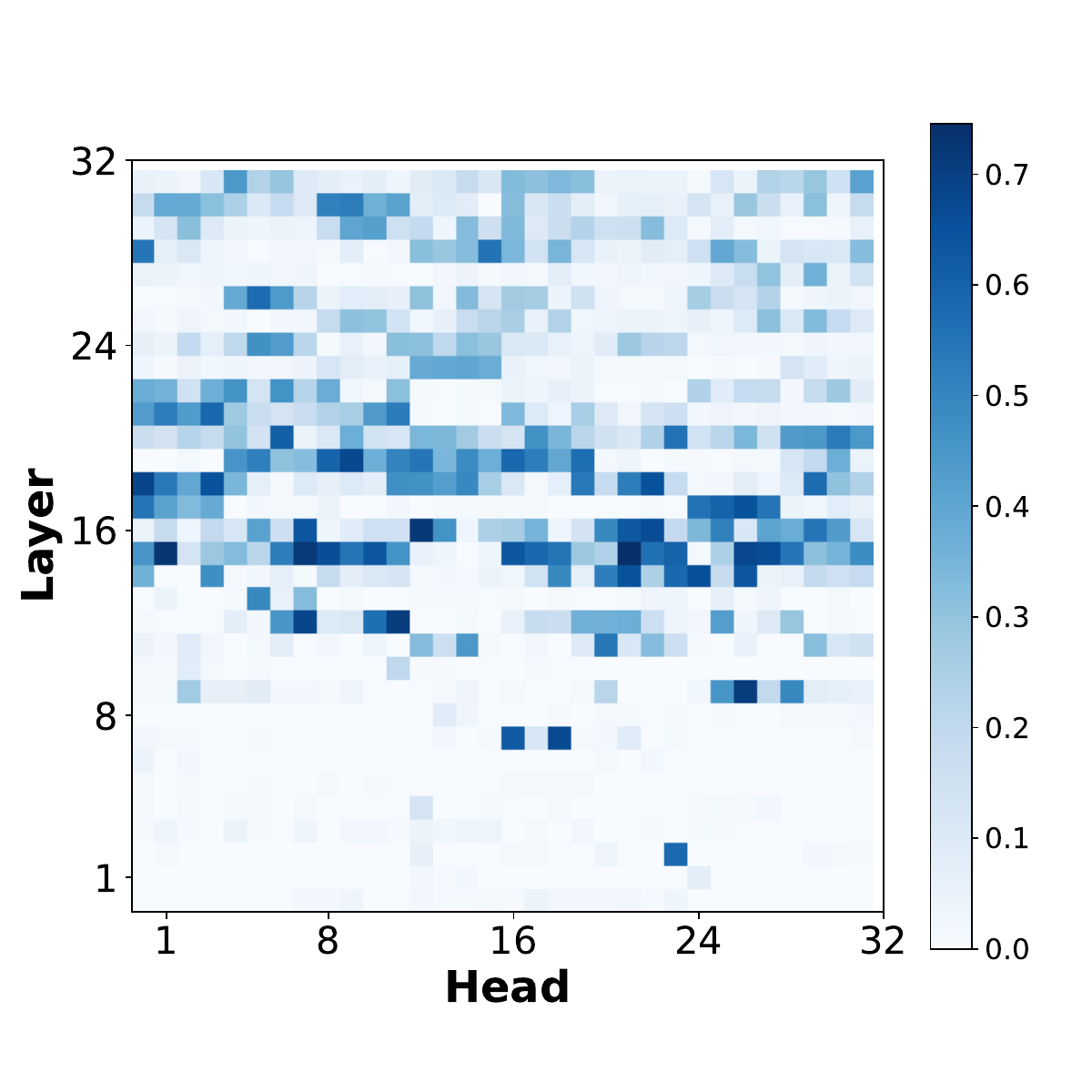}
    \end{minipage}
    \begin{minipage}{0.49\linewidth}
        \includegraphics[scale=0.17,trim={0 0.5cm 0 0.5cm},clip]{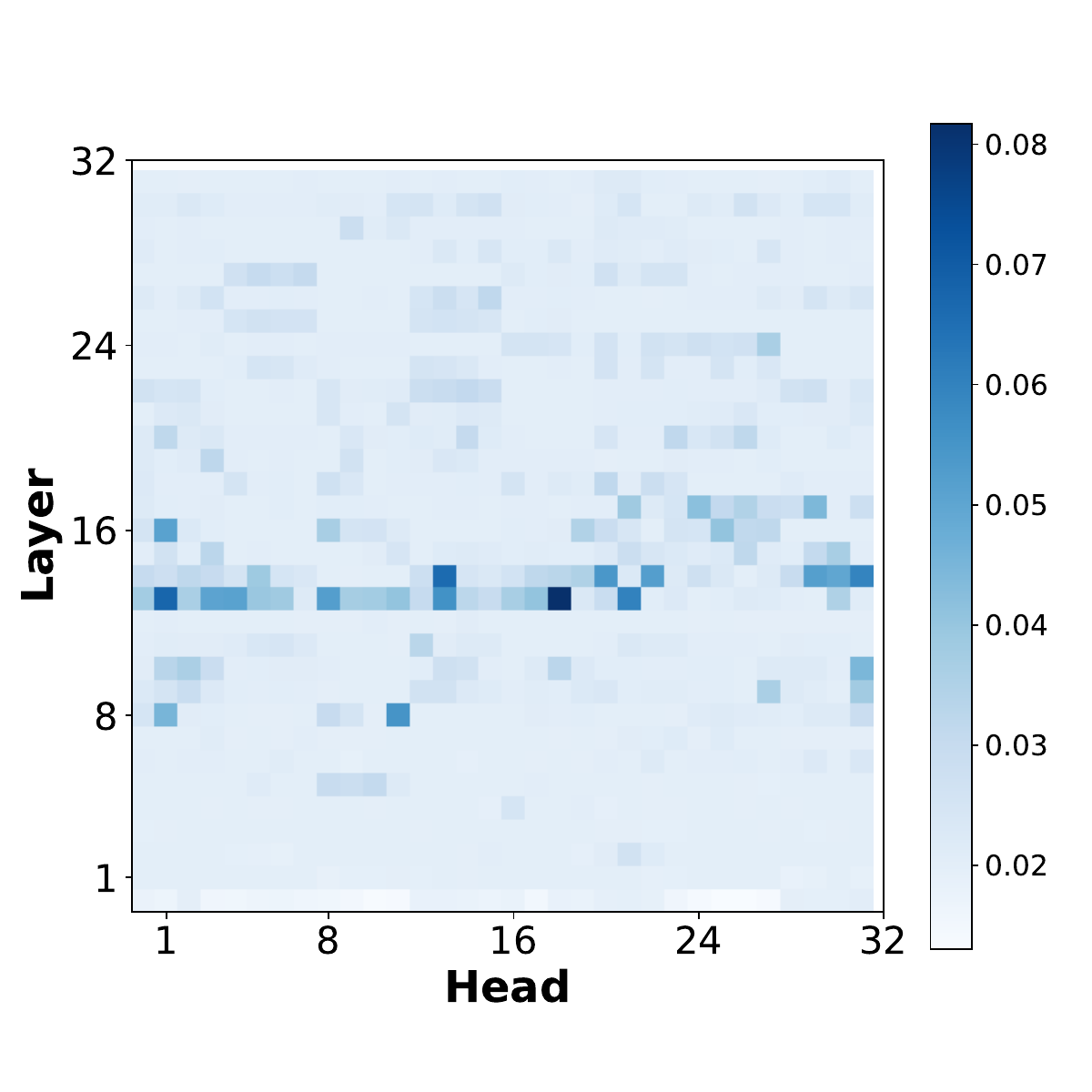}
    \end{minipage}
  \caption{Distribution of $S_{CoRe}$ in Mistral (left) and Llama (right).}\label{fig.heatmap}
\end{wrapfigure}

\textbf{CoRe heads consistently localize in middle layers. }
Figure~\ref{fig.heatmap} shows the distribution of $S_{CoRe}$ across all attention heads in Mistral ($t=0.001$) and Llama ($t=0.1$). Although the score distributions differ across models due to different temperature settings, both exhibit a strikingly consistent pattern: the highest-scoring CoRe heads are concentrated in middle transformer layers.

Importantly, this localization pattern remains highly stable across different detection datasets (Appendix~\ref{app.msmarco}), different hard-negative constructions (lexical vs.\ semantic), and different positions of the positive document within the candidate set, e.g., early, middle, late, random (Appendix~\ref{app.negative.exp}). Temperature controls the sharpness of contrastive discrimination; empirically, however, both CoRe head localization and reranking performance remain stable across a broad temperature range (Appendix~\ref{app.temperature}).

These findings suggest that retrieval-relevant computation in transformer rerankers is structurally localized rather than uniformly distributed across layers. This observation is also consistent with prior studies on transformer representations, which suggest that middle layers capture richer semantic information while later layers are more closely associated with token generation \citep{clark2019does, van2019does}.

\textbf{Middle-layer localization enables efficient reranking via layer pruning. }
The strong concentration of CoRe heads in middle transformer layers suggests that retrieval-relevant computation is structurally localized and partially separable from later-stage generative processing. To test this hypothesis, we progressively prune upper transformer layers during inference and evaluate the resulting accuracy--efficiency tradeoff.

\begin{figure*}[t]
\begin{minipage}[t]{0.5\textwidth}
  \includegraphics[scale=0.27]{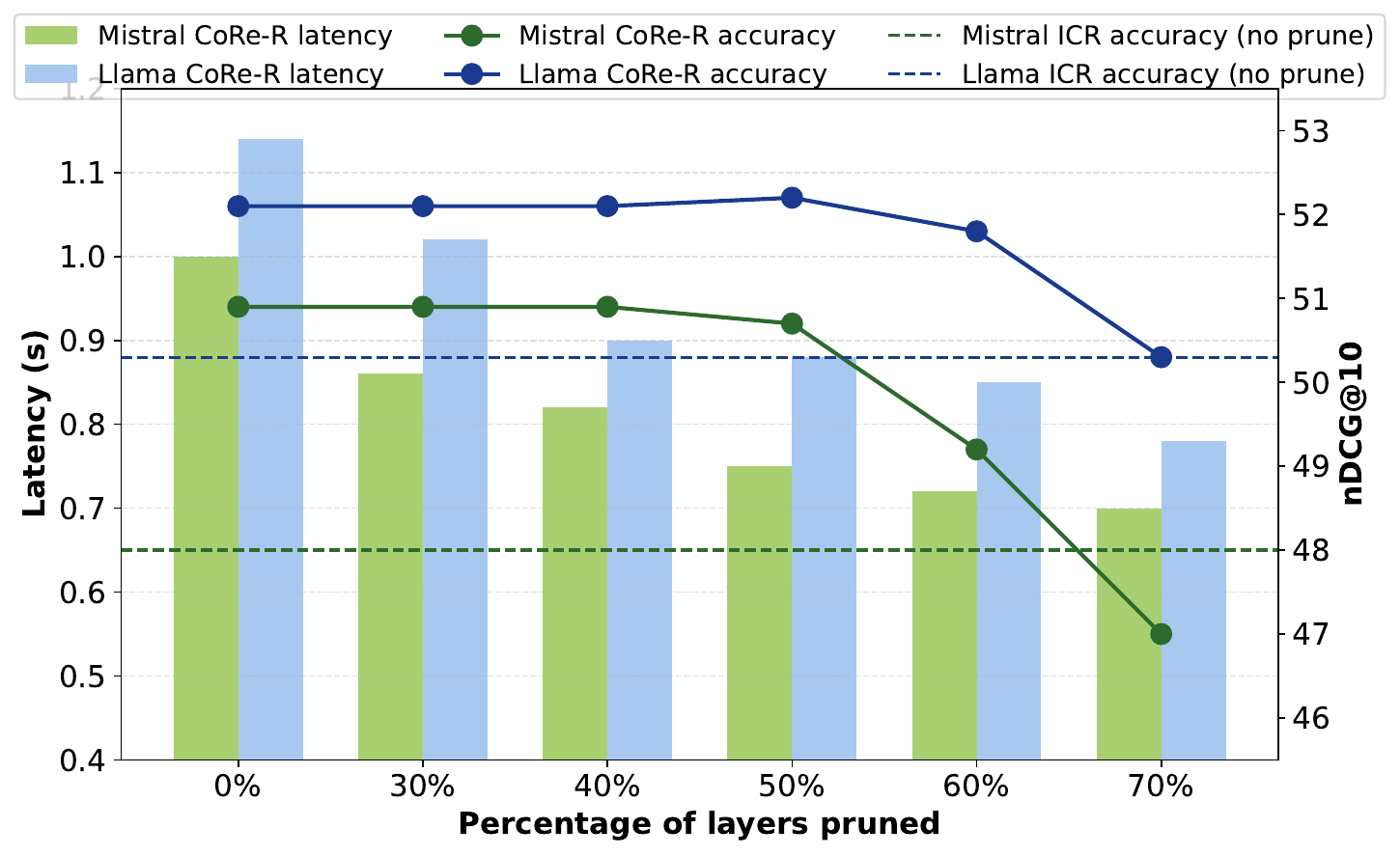}
  \caption{Average accuracy and latency on BEIR.}\label{fig.runtime}
\end{minipage}%
\hspace{5pt}
\begin{minipage}[t]{0.5\textwidth}
  \includegraphics[scale=0.27]{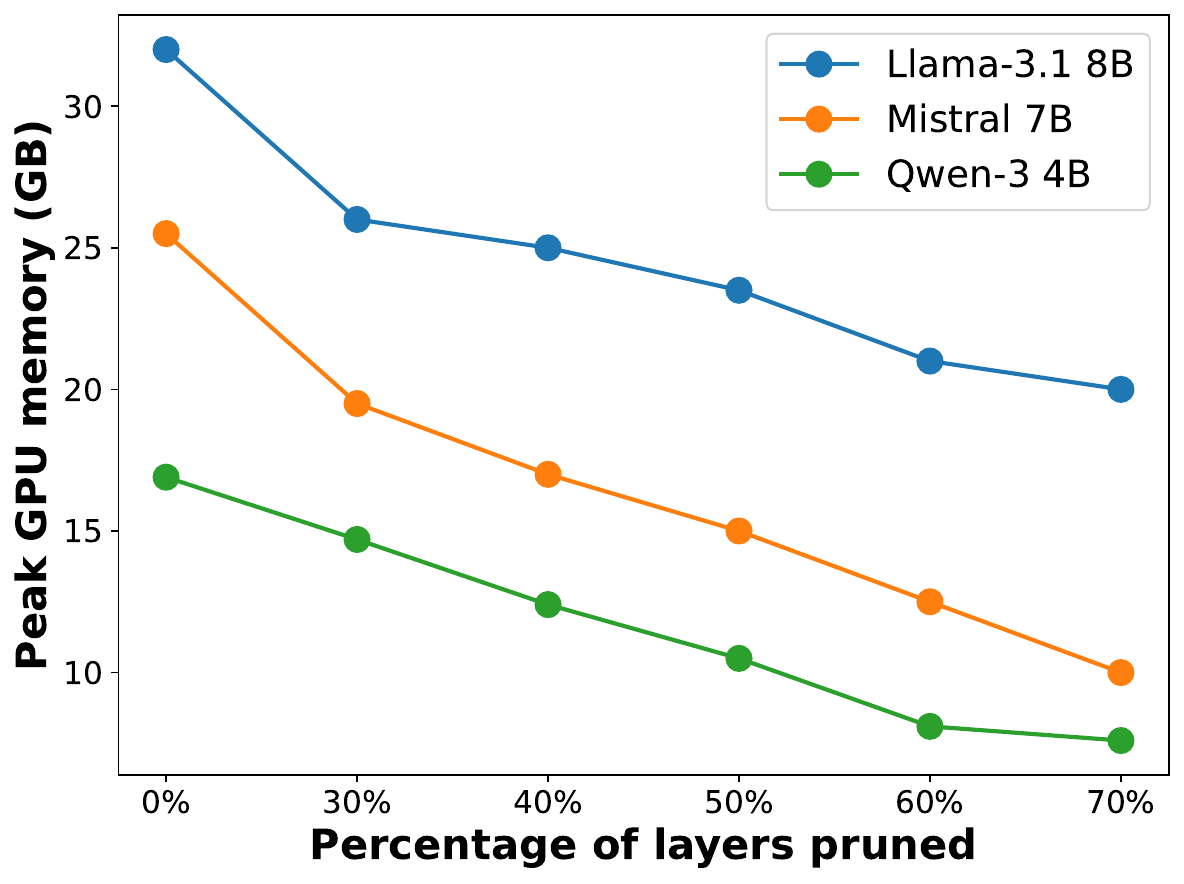}
    \caption{Peak GPU memory usage on BEIR.}\label{fig.memory}
\end{minipage}%
\end{figure*}

Figure~\ref{fig.runtime} reports reranking accuracy and latency on BEIR under different pruning levels, while Figure~\ref{fig.memory} shows the corresponding peak GPU memory usage measured on a single H100 96GB GPU. Despite pruning the final $50\%$ of transformer layers in both Mistral and Llama, CoRe-R preserves nearly identical reranking performance while reducing GPU memory usage by approximately $40\%$ and inference latency by $20\%$. Similar trends are observed for Qwen (Appendix~\ref{app.additional.pruning}).

In contrast, retrieval-head baselines including QR heads and NIAH heads~\citep{wu2025retrieval} remain stable only under moderate pruning ($30$--$40\%$) before degrading substantially (Appendix~\ref{app.baseline.prune}). Even under aggressive $60\%$ pruning, CoRe-R continues to outperform the ICR baseline across all evaluated models. These results suggest that reranking-relevant computation in transformer rerankers is substantially more sparse and localized than previously assumed. In particular, a small subset of middle-layer CoRe heads appears to capture most retrieval-relevant information, while many later transformer layers contribute comparatively little to reranking quality.

\begin{wrapfigure}[13]{r}{0.41\textwidth}
\vspace{-0.6cm}
  \begin{center}
    \includegraphics[scale=0.27]{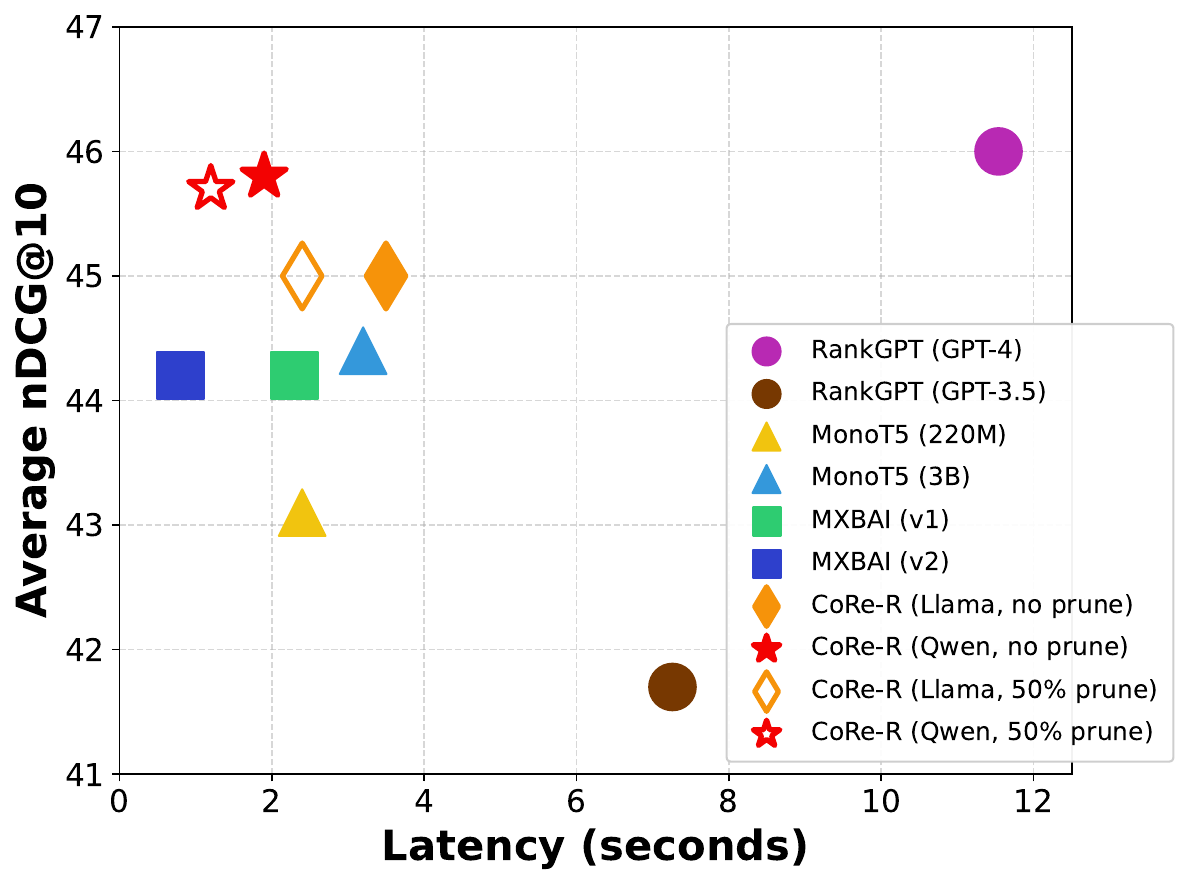}
  \caption{Accuracy vs. latency on a subset of BEIR top-40.}\label{fig.tradeoff}
  \end{center}
\end{wrapfigure}

\textbf{Comparison of reranking quality and inference efficiency.}
To contextualize the practical quality-efficiency tradeoff of CoRe-R, we compare against two representative reranking paradigms: the generative zero-shot list-wise reranker RankGPT~\citep{sun2023chatgpt} and the supervised point-wise reranker MonoT5~\citep{nogueira2020document} and MXBAI~\cite{li2025prorank}. Figure~\ref{fig.tradeoff} reports averaged reranking accuracy on a BEIR subset together with the corresponding inference latency (see Appendix~\ref{app.point.list} for detailed results).

CoRe-R lies on the accuracy-latency Pareto frontier across all the evaluated reranking methods. CoRe-R based on both Llama and Qwen outperforms RankGPT-3.5, while CoRe-R (Qwen) remains competitive with RankGPT-4 despite substantially lower latency. Compared to MonoT5 and MXBAI, CoRe-R achieves stronger reranking performance under similar latency budgets, and becomes even faster when layer pruning is applied.

\section{Conclusion}

We introduced \emph{Contrastive Retrieval heads} (CoRe heads), a subset of attention heads that capture the most discriminative signals for document reranking. We proposed a contrastive scoring metric that rewards attention to relevant documents while penalizing attention to irrelevant ones, enabling reliable identification of informative heads. Across extensive experiments, aggregating attention from CoRe heads yields a state-of-the-art reranker that consistently outperforms prior baselines across tasks and models.

Our analysis further shows that CoRe heads concentrate in the middle transformer layers, enabling effective layer pruning that reduces inference latency and memory usage with minimal impact on accuracy. Consistent with our theoretical analysis, CoRe also produces larger empirical gaps between consecutively ranked head scores than QR, indicating sharper separation between informative and non-informative heads. Together, these findings suggest that CoRe heads are primary carriers of relevance information and provide a practical foundation for fast, accurate, and efficient retrieval systems.

\textbf{Limitations.} Attention-based rerankers can be sensitive to task-specific instruction design, particularly on non-QA retrieval tasks such as Quora, Climate-FEVER, and ArguAna. Our experiments primarily use a universal instruction prompt shared across datasets, which may not fully align with dataset-specific retrieval objectives such as duplicate detection or counter-evidence retrieval. As shown in Appendix~\ref{app.result.prompt}, customized prompts can substantially improve reranking performance and allow CoRe-R to outperform the retrieval baseline. Future work could explore more systematic prompt adaptation strategies for different retrieval settings.

\section*{Acknowledgements}
This work is funded in part through the IBM-RPI Future of Computing Research Collaboration.

\bibliography{references}
\bibliographystyle{plain}

\newpage
\appendix
\onecolumn

\section{Analysis of contrastive head selection}

\subsection{Proof of Proposition~\ref{thm:gap} }
\begin{proof}
For any fixed head $h\in\mathcal{H}$, Hoeffding's inequality gives
\[
\Pr\left(
|\hat{\mu}_h-\mu_h|>\frac{\Delta}{2}
\right)
\leq
2\exp\left(-2N\left(\frac{\Delta}{2}\right)^2\right)
=
2\exp\left(-\frac{N\Delta^2}{2}\right).
\]
Applying a union bound over all heads in $\mathcal{H}$, we obtain
\[
\Pr\left(
\max_{h\in\mathcal{H}}|\hat{\mu}_h-\mu_h|
\leq
\frac{\Delta}{2}
\right)
\geq
1-
2|\mathcal{H}|
\exp\left(-\frac{N\Delta^2}{2}\right).
\]

Now consider the event
\[
\mathcal{E}
=
\left\{
\max_{h\in\mathcal{H}}|\hat{\mu}_h-\mu_h|
\leq
\frac{\Delta}{2}
\right\}.
\]
On this event, for any $h\in A^*$ and any $h'\notin A^*$, the definition of the gap implies
\[
\mu_h-\mu_{h'}\geq \Delta.
\]
Therefore,
\[
\hat{\mu}_h
\geq
\mu_h-\frac{\Delta}{2}
\geq
\mu_{h'}+\frac{\Delta}{2}
\geq
\hat{\mu}_{h'}.
\]
Thus every head in $A^*$ has empirical score at least as large as every head outside $A^*$. With any fixed tie-breaking rule, the empirical top-$m$ set equals the population top-$m$ set, i.e.,
\[
\hat{A}=A^*.
\]
Since the event $\mathcal{E}$ holds with probability at least
\[
1-
2|\mathcal{H}|
\exp\left(-\frac{N\Delta^2}{2}\right),
\]
the result follows.
\end{proof}

\subsection{Proof of Proposition~\ref{prop:qr_suboptimal}}
\label{app.proof.prop}

\begin{proof}
Consider the case with $k=2$ documents: one positive document $d^+$ and one negative document $d^-$.
Let the document-level scores satisfy
\[
s^a(d^+) > s^b(d^+)
\quad\text{and}\quad
s^a(d^-) \gg s^a(d^+),
\]
while
\[
s^b(d^+) > s^b(d^-).
\]
Such configurations are feasible when document-level scores do not preserve the normalization post-softmax. In such cases, head $b$ is more preferable than head $a$. 

Under QR, head $a$ is ranked above head $b$ since it assigns a larger score to the positive document.
However, the contrastive score
\[
S_{CoRe}(h) = \frac{\exp(s^h(d^+)/t)}{\exp(s^h(d^+)/t) + \exp(s^h(d^-)/t)}
\]
is decreasing in the margin $s^h(d^-)-s^h(d^+)$.
Therefore, head $a$, which assigns more attention to the negative document than to the positive, satisfies $S_{CoRe}(a) < S_{CoRe}(b)$.

The same reasoning applies to any $k>2$, as $S_{CoRe}(h)$ depends monotonically on the total attention assigned to negative documents.
Thus, QR may select heads with inferior positive--negative separation even in larger candidate sets.
\end{proof}

\section{Experiment details}

\subsection{Reranking configuration}
\label{app.config}

We conduct the reranking experiments using a variety of base retriever. For the standard BEIR benchmark, we use a lightweight dense retriever \textit{ibm-granite/granite-embedding-30m-english}. For the multilingual benchmark MLDR, we use the multilingual version \textit{ibm-granite/granite-embedding-107m-multilingual}. We use BM25 for BRIGHT and ColBERT-v2 for multi-hop datasets following \cite{chenattention}. We also report experiment using more modern retriever \textit{Qwen/Qwen3-Embedding-0.6B} \citep{qwen3embedding} in Appendix~\ref{app.qwen.retriever}. In all settings, we rerank top-40 documents in all settings using a single GPU H100 96GB.

\subsection{Prompt structure}
\label{app.prompt}

We use the same prompt structure for both head detection and reranking steps, which is adopted from \citep{chenattention}. The prompt starts with the instruction \texttt{Here are some paragraphs:} and the list of retrieved documents separated by a newline. After the list of document, there is a second prefix prompt for the query \texttt{Please find information that are relevant to the following query in the paragraphs above. Query:}, and the query is appended at the end. The entire prompt string is wrapped by the special tokens \texttt{start token} and \texttt{end token} which vary based on the LLM used. See Table~\ref{app.table.prompt} for an example.

\begin{table}[h]
    \centering
    \caption{An example prompt from NQ with Mistral 7B.}
    \begin{tabular}{|l|}
    \hline
    \texttt{[INST] Here are some paragraphs:} \\
    \\
    \texttt{[document 1]} Can't Help Falling in Love``Can't Help Falling in Love" is a pop ballad \\
    originally recorded by American singer Elvis Presley and published by Gladys Music, \\
    Presley's publishing company. It was written by Hugo Peretti, Luigi Creatore, and \\
    George David Weiss. The song was featured in Presley's 1961 film, Blue Hawaii. \\
    \\
    \texttt{[document 2]} Can't Help Falling in Love In 2015, the song was included on the If I \\
    Can Dream album, on the occasion of the 80th anniversary of Presley's birth. The version \\
    uses archival voice recordings of Presley and his singers, backed by new orchestral \\
    arrangements performed by the Royal Philharmonic Orchestra. \\
    \\
    ... \\
    \\
    \texttt{[document 40]} I Can't Help It (If I'm Still in Love with You) Williams sang the song \\
    with Anita Carter on the Kate Smith Evening Hour on April 23, 1952. The rare television \\
    appearance is one of the few film clips of Williams in performance. \\
    \\
    \texttt{Please find information that are relevant to the following} \\
    \texttt{query in the paragraphs above.} \\
    \\
    \texttt{Query:} who recorded i can't help falling in love with you\texttt{[/INST]} \\
    \hline
    \end{tabular}
    \label{app.table.prompt}
\end{table}

\section{Distribution of retrieval heads}

\subsection{Head location}
\label{app.head.location}

Prior work \citep{wu2025retrieval} show that only the top $10-15$ heads demonstrate strong retrieval signals. Our empirical results on the optimal number of heads (Section~\ref{app.optimal.core.head}) support this observation, as the reranking accuracy peaks within the top-$10$ heads and starts to decline after $10$ heads. This indicates the top few heads are the most important heads for retrieval tasks, while heads outside of top-$10$ are less effective which may contribute noise to the retrieval process.

We discuss the overlapping between different set of retrieval heads and compare their performance on final re-ranking tasks. For Llama, none of the top $8$ CoRe heads overlaps with the top $8$ NIAH heads, and only $3$ overlap with the top $8$ QR heads. For Mistral, top $8$ CoRe heads overlap with $2$ QR heads and $2$ NIAH heads. In addition, CoRe heads concentrate in lower layers than both QR heads and NIAH heads.

We list below the exact location of the top $8$ NIAH retrieval heads, QR heads and CoRe heads for Llama-3.1 8B and Mistral 7B. We format ($L$-$H$) as layer $L$ head $H$, for example, ($7$-$19$) means layer $7$ head $19$.

Llama-3.1 8B:
\begin{itemize}
    \item NIAH retrieval heads: ($15$-$30$), ($27$-$7$), ($8$-$1$), ($16$-$1$), ($24$-$27$), ($16$-$20$), ($5$-$8$), ($16$-$23$).
    \item QR heads: ($13$-$18$), ($14$-$13$), ($13$-$1$), ($20$-$14$), ($14$-$29$), ($16$-$1$), ($14$-$22$), ($17$-$29$).
    \item CoRe heads: ($13$-$18$), ($13$-$1$), ($14$-$13$), ($13$-$21$), ($14$-$31$), ($13$-$13$), ($8$-$11$), ($14$-$20$).
\end{itemize}

Mistral 7B:
\begin{itemize}
    \item NIAH retrieval heads: ($18$-$0$), ($12$-$7$), ($12$-$6$), ($18$-$2$), ($18$-$3$), ($18$-$1$), ($30$-$8$), ($28$-$0$).
    \item QR heads: ($18$-$22$), ($15$-$26$), ($20$-$17$), ($18$-$0$), ($19$-$9$), ($16$-$22$), ($16$-$12$), ($19$-$16$).
    \item CoRe heads: ($15$-$21$), ($15$-$1$), ($16$-$12$), ($15$-$7$), ($9$-$26$), ($12$-$11$), ($12$-$7$), ($18$-$0$).
\end{itemize}

We found that for Mistral 7B, CoRe heads overlap in $2$ heads with both QR heads and NIAH retrieval heads. For Llama-3.1 8B, our top CoRe heads overlap with QR heads in $3$ heads, but do not overlap with NIAH retrieval heads. Notably, the common 3 heads of CoRe and QR are in the top-3 and these are the heads with highest retrieval score. This explains why there is a close gap between QR-R and CoRe-R in the Llama results. Nevertheless, due to the difference in the remaining 5 heads, CoRe-R shows superior performance in many datasets that exhibit high number of hard negatives. In both models, CoRe heads appear in earlier layers compared to other heads, allowing for better accuracy-efficiency tradeoff with layer pruning.

We also investigate the overlapping in top-32 and top-64 heads between CoRe heads and the baselines in Table~\ref{app.table.overlap}. In both top-32 and top-64 heads, we found that CoRe heads overlaps in approximately 50\% with NIAH and 75\% with QR. However, as mentioned above, these lower-ranking heads are not important for retrieval tasks, and their retrieval scores are very low compared to the first top-8 heads.

\begin{table}[h]
    \centering
    \begin{tabular}{|c|c|c|c|c|}
        \hline
        Baselines & Mistral Top-32 & Mistral Top-64 & Llama Top-32 & Llama Top-64 \\
        \hline\hline
        NIAH heads & 19 & 35 & 16 & 30 \\
        QR heads & 25 & 48 & 27 & 53 \\
        \hline
    \end{tabular}
    \caption{Number of overlapping heads between CoRe heads and other baselines.}
    \label{app.table.overlap}
\end{table}

\subsection{Empirical score separation of retrieval heads}
\label{app.head.score.gap}

\begin{figure*}[h]
\centering
\begin{subfigure}{0.47\textwidth}
\centering
  \includegraphics[width=0.9\textwidth]{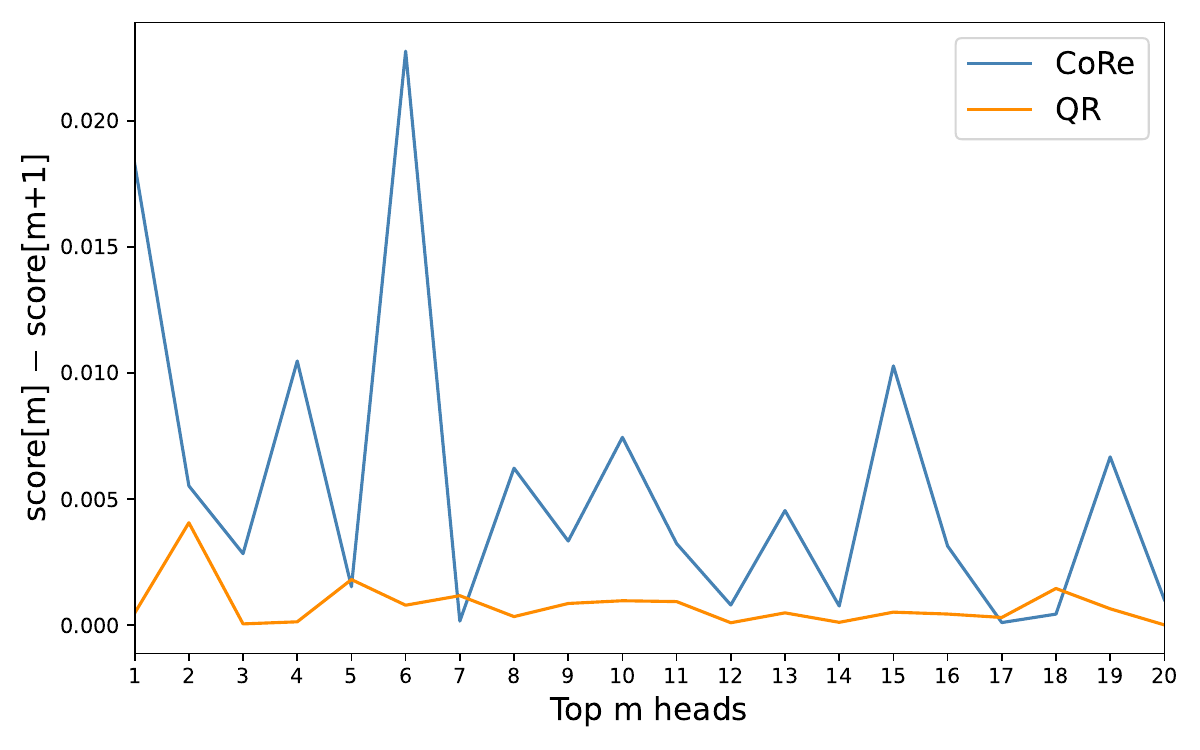}
  \caption{Mistral 7B}
  \label{fig.score.gap.mistral}
\end{subfigure}
\begin{subfigure}{0.47\textwidth}
\centering
  \includegraphics[width=0.9\textwidth]{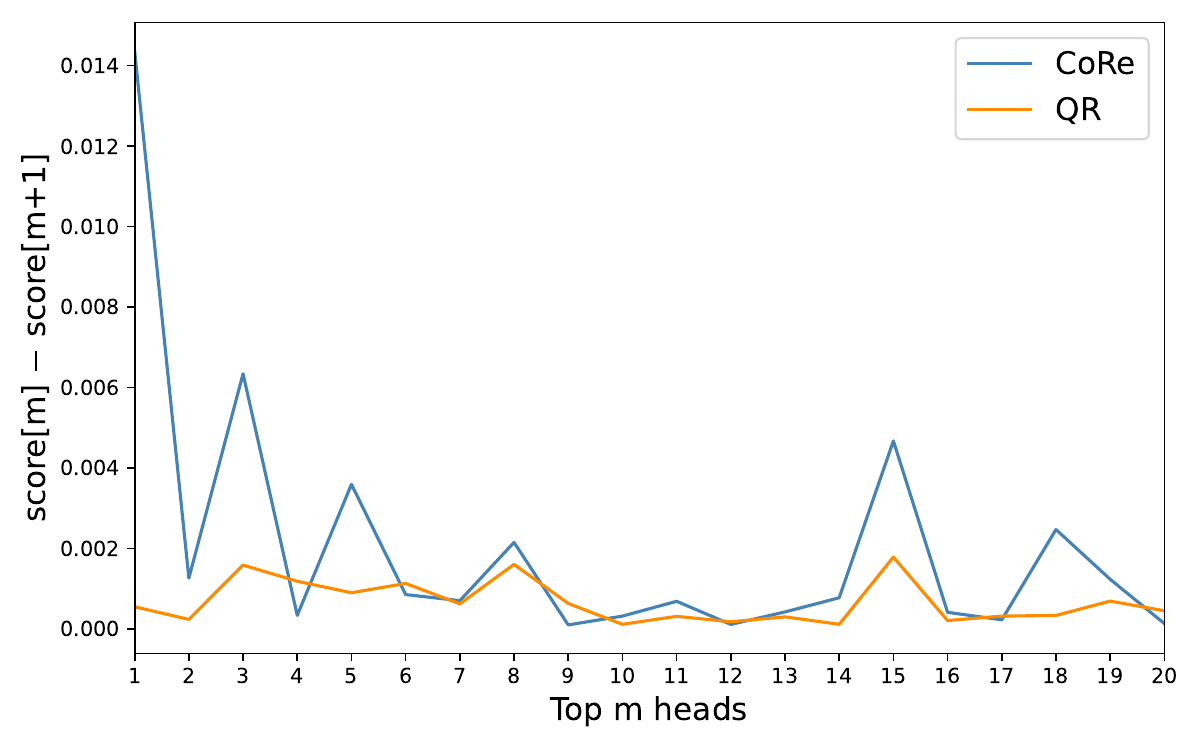}
  \caption{Llama-3.1 8B}
  \label{fig.score.gap.llama}
\end{subfigure}
\caption{Adjacent score gaps between consecutively ranked CoRe and QR heads.}
\label{fig.score.gap}
\end{figure*}

Figure~\ref{fig.score.gap} compares the adjacent score gaps between consecutively ranked CoRe and QR heads. Across both Llama and Mistral, CoRe exhibits substantially sharper separation among top-ranked heads, whereas QR produces a considerably flatter score distribution. This indicates that CoRe concentrates retrieval-relevant signals into a smaller set of highly discriminative heads, while QR yields a more diffuse ranking over attention heads. Consistent with the gap condition in Theorem~\ref{thm:gap}, the stronger empirical separation induced by CoRe helps explain the stability and reproducibility of the detected CoRe head sets across datasets and hard-negative configurations.

\section{Additional experiment results}

\subsection{Comparison of rerankers using NIAH heads, QR heads and CoRe heads}
\label{app.baseline.main}

Table~\ref{table.niah.and.qr} compares CoRe-R with the attention-based rerankers using the NIAH heads (NIAH-R) for Llama. As NIAH retrieval heads are limited to copy-paste operations, NIAH-R underperforms ICR on average, showing gains only on a few datasets such as Climate-FEVER and ArguAna, while showing worse performance on many other datasets. Across all datasets, CoRe-R shows the highest overall nDCG@10 compared to all baselines.

\setlength{\tabcolsep}{3pt}
\begin{table}[h]
\caption{nDCG@10 on BEIR top-$40$ with Llama.}
\label{table.niah.and.qr}
\scriptsize
\begin{center}
\centering
\begin{tabular}{ | c | c | c c c c | }
\hline
Task & Dense & ICR & NIAH-R & QR-R & CoRe-R \\
& baseline & & & & \\
\hline\hline
TREC-COVID     & 63.1 & \textbf{76.8} & 73.7 & 75.7 & 75.7 \\
NFCorpus       & 33.7 & 35.0 & 34.2 & 36.4 & \textbf{36.7} \\
DBPedia        & 36.0 & 38.7 & 37.1 & 39.0 & \textbf{39.4} \\
SciFact        & 71.3 & 74.7 & 74.2 & \textbf{75.9} & 75.1 \\
SciDocs        & 22.5 & 18.5 & 19.2 & 19.7 & \textbf{20.2} \\
FiQA           & 36.9 & 41.3 & 40.8 & 44.0 & \textbf{44.1} \\
NQ             & 51.6 & 60.8 & 57.5 & 63.0 & \textbf{63.2} \\
FEVER          & 85.5 & \textbf{89.2} & 88.1 & 87.4 & 88.4 \\
Climate-FEVER  & 30.3 & 22.5 & \textbf{23.1} & 22.9 & 22.7 \\
HotpotQA       & 62.9 & 73.7 & 72.0 & 73.6 & \textbf{73.8} \\
Touche         & 24.0 & 26.1 & 25.5 & 26.3 & \textbf{26.4} \\
MSMARCO        & 30.7 & 32.9 & 31.6 & 33.9 & \textbf{34.9} \\
Quora          & 86.7 & 78.7 & 80.1 & 74.6 & \textbf{82.3} \\
ArguAna        & 56.4 & 42.5 & 51.7 & 54.2 & \textbf{55.0} \\
CQADupstack    & 44.3 & 41.7 & 41.4 & 43.2 & \textbf{43.9} \\
\hline
Average        & 49.1 & 50.2 & 50.0 & 51.3 & \textbf{52.1} \\
\hline
\end{tabular}
\end{center}
\end{table}

\subsection{Effect of candidate document ordering}
\label{app.doc.order}

Table~\ref{table.order} shows the reranking accuracy for various orders of the documents, including the original order retrieved by the base retriever, the reversed order and random order (with seed $42$). We generally found that using the original order leads to the best performance. While the absolute performance varies under different orderings, CoRe-R consistently outperforms ICR and QR-R across all settings. This suggests that CoRe head selection is not driven primarily by positional heuristics, but captures more robust contrastive signals between relevant and non-relevant documents.

\setlength{\tabcolsep}{4pt}
\begin{table}[h]
\caption{nDCG@10 on the BEIR benchmark top-$40$ using Llama with different document orders.}
\label{table.order}
\begin{center}
\scriptsize
\begin{tabular}{ | c | c | c c c | c c c | c c c | }
\hline
Task & Dense & \multicolumn{3}{c|}{Original order} & \multicolumn{3}{c|}{Reversed order} & \multicolumn{3}{c|}{Random order} \\\cline{3-11}
& baseline & ICR & QR-R & CoRe-R & ICR & QR-R & CoRe-R & ICR & QR-R & CoRe-R \\
\hline\hline
TREC-COVID & 63.1 & \textbf{76.8} & 75.7 & 75.7     & 73.2 & \textbf{74.5} & 74.4       & 70.7 & 71.5 & \textbf{73.4} \\
NFCorpus & 33.7 & 35.0 & 36.4 & \textbf{36.7}       & 35.1 & 32.4 & \textbf{36.4}       & 33.8 & 35.5 & \textbf{35.9} \\
DBPedia & 36.0 & 38.7 & 39.0 & \textbf{39.4}        & 36.8 & 35.8 & \textbf{37.1}       & 35.6 & 35.9 & \textbf{36.6} \\
SciFact & 71.3 & 74.7 & \textbf{75.9} & 75.1        & 75.7 & 75.4 & \textbf{76.1}       & 74.8 & 75.7 & \textbf{76.2} \\
SciDocs & 22.5 & 18.5 & 19.7 & \textbf{20.2}        & 19.9 & 20.8 & \textbf{21.7}       & 17.9 & 19.3 & \textbf{20.2} \\
FiQA & 36.9 & 41.3 & 44.0 & \textbf{44.1}           & 39.8 & 40.1 & \textbf{43.0}       & 39.3 & 40.6 & \textbf{42.7} \\
NQ & 51.6 & 60.8 & 63.0 & \textbf{63.2}             & 56.6 & 60.9 & \textbf{62.1}       & 58.4 & 61.9 & \textbf{62.8} \\
FEVER & 85.5 & \textbf{89.2} & 87.4 & 88.4          & 87.9 & 88.1 & \textbf{88.9}       & \textbf{87.0} & 86.5 & 86.2 \\
Climate-FEVER & 30.3 & 22.5 & \textbf{22.9} & 22.7  & 23.1 & 23.3 & \textbf{24.6}       & 21.5 & \textbf{22.2} & 21.9 \\
HotpotQA & 62.9 & 73.7 & 73.6 & \textbf{73.8}       & 72.9 & 73.1 & \textbf{74.0}       & 72.6 & 72.8 & \textbf{73.3} \\
Touche & 24.0 & 26.1 & 26.3 & \textbf{26.4}         & 26.6 & 27.2 & \textbf{27.5}       & 24.5 & 25.6 & \textbf{26.8} \\
MSMARCO & 30.7 & 32.9 & 33.9 & \textbf{34.9}        & 28.7 & 29.7 & \textbf{30.4}       & 30.6 & 31.8 & \textbf{32.4} \\
Quora & 86.7 & 78.7 & 74.6 & \textbf{82.3}          & 62.3 & 64.0 & \textbf{79.2}       & 73.2 & 74.8 & \textbf{80.8} \\
ArguAna & 56.4 & 42.5 & 54.2 & \textbf{55.0}        & 41.6 & 51.3 & \textbf{53.9}       & 40.3 & 52.4 & \textbf{54.3} \\
CQADupstack & 44.3 & 41.7 & 43.2 & \textbf{43.9}    & 40.6 & 41.5 & \textbf{42.8}       & 41.4 & 42.2 & \textbf{43.5} \\
\hline
Average & 49.1 & 50.2 & 51.3 & \textbf{52.1}        & 48.1 & 49.2 & \textbf{51.5}       & 48.1 & 49.9 & \textbf{51.1} \\
\hline
\end{tabular}
\end{center}
\end{table}


\subsection{Results using BM25 retriever on BEIR benchmark.}
\label{app.beir.bm25}

In addition to the main results using dense retriever in Table~\ref{table.beir}, we repeat the reranking experiment from using BM25 and report the average score on BEIR in Table~\ref{table.beir.bm25.top40}. CoRe-R remains the strongest attention-based reranker for both models, improving over ICR by 3.9–5.1 nDCG and over QR-R by 1.4–2.2 nDCG. This confirms that the gains are not specific to candidates produced by a dense retriever.

\setlength{\tabcolsep}{4pt}
\begin{table}[h]
\caption{nDCG@10 on the BEIR benchmark top-$40$ with BM25.}
\label{table.beir.bm25.top40}
\begin{center}
\scriptsize
\begin{tabular}{ | c | c | c c c | c c c | c c c | c c c | }
\hline
Dataset & Retriever & \multicolumn{3}{c|}{Mistral 7B} & \multicolumn{3}{c|}{Llama-3.1 8B} & \multicolumn{3}{c|}{Phi-4} & \multicolumn{3}{c|}{Granite-3.2 8B} \\\cline{3-14}
& baseline & ICR & QR-R & CoRe-R & ICR & QR-R & CoRe-R & ICR & QR-R & CoRe-R & ICR & QR-R & CoRe-R \\
\hline\hline
TREC-COVID & 0 & 59.7 & 63.2 & \textbf{64.3} & \textbf{76.8} & 75.7 & 75.7 & 75.2 & \textbf{76.8} & 76.0 & 70.4 & 70.3 & \textbf{71.8} \\
NFCorpus & 0 & 32.7 & 34.5 & \textbf{35.0} & 35.0 & 36.4 & \textbf{36.7} & 31.6 & 35.5 & \textbf{35.6} & 31.8 & 32.5 & \textbf{33.9} \\
DBPedia & 0 & 32.2 & 33.4 & \textbf{34.7} & 38.7 & 39.0 & \textbf{39.4} & 36.9 & 39.4 & \textbf{39.6} & 35.6 & 35.9 & \textbf{36.8} \\
SciFact & 0 & 68.8 & 68.1 & \textbf{74.7} & 74.7 & \textbf{75.9} & 75.1 & 71.7 & 74.2 & \textbf{74.7} & 74.6 & 74.7 & \textbf{75.2} \\
SciDocs & 0 & 16.6 & 17.3 & \textbf{19.5} & 18.5 & 19.7 & \textbf{20.2} & 17.8 & 21.0 & \textbf{21.1} & 18.4 & \textbf{20.3} & 20.0 \\
FiQA & 0 & 35.8 & 40.3 & \textbf{41.4} & 41.3 & 44.0 & \textbf{44.1} & 37.2 & 43.8 & \textbf{44.1} & 39.7 & 41.6 & \textbf{42.4} \\
NQ & 0 & 55.4 & 56.5 & \textbf{58.1} & 60.8 & 63.0 & \textbf{63.2} & 57.4 & 63.3 & \textbf{63.6} & 57.4 & 57.9 & \textbf{60.7} \\
FEVER & 0 & 87.5 & 87.1 & \textbf{88.7} & \textbf{89.2} & 87.4 & 88.4 & 89.1 & \textbf{89.6} & \textbf{89.6} & \textbf{88.0} & 87.5 & 86.2 \\
Climate-FEVER & 0 & 21.6 & 21.1 & \textbf{23.3} & 22.5 & \textbf{22.9} & 22.7 & 22.2 & \textbf{24.5} & 24.3 & 20.5 & \textbf{21.2} & 21.0 \\
HotpotQA & 0 & 71.7 & 71.6 & \textbf{73.4} & 73.7 & 73.6 & \textbf{73.8} & 73.3 & 74.2 & \textbf{74.5} & \textbf{73.6} & 72.8 & 72.6 \\
Touche & 0 & 23.3 & 26.4 & \textbf{26.7} & 26.1 & 26.3 & \textbf{26.4} & 21.8 & 25.4 & \textbf{25.7} & 20.5 & 24.6 & \textbf{24.8} \\
MSMARCO & 0 & 30.1 & \textbf{32.1} & 31.7 & 32.9 & 33.9 & \textbf{34.9} & 30.6 & 34.6 & \textbf{34.8} & 29.6 & 31.2 & \textbf{32.5} \\
Quora & 0 & 81.1 & 71.5 & \textbf{85.2} & 78.7 & 74.6 & \textbf{82.3} & 74.2 & 76.5 & \textbf{80.2} & \textbf{83.1} & 79.8 & 75.1 \\
ArguAna & 0 & 45.0 & 51.1 & \textbf{52.7} & 42.5 & 54.2 & \textbf{55.0} & 41.7 & 52.2 & \textbf{52.7} & 52.3 & 56.1 & \textbf{57.3} \\
CQADupstack & 0 & 38.7 & 40.8 & \textbf{41.6} & 41.7 & 43.2 & \textbf{43.9} & 39.4 & 44.0 & \textbf{44.3} & 41.4 & 42.3 & \textbf{43.4} \\
\hline
Average & 40.8 & 47.9 & 48.9 & \textbf{50.9} & 50.2 & 51.3 & \textbf{52.1} & 48.0 & 51.7 & \textbf{52.1} & 49.1 & 50.0 & \textbf{50.3} \\
\hline
\end{tabular}
\end{center}
\end{table}

\subsection{Results under varying context lengths}
\label{app.long.context}

In this subsection, we evaluate the attention-based rerankers on various context length with the BEIR benchmark. We report per-dataset nDCG@10 using top-$60$ documents in Table~\ref{table.top60}, top-$80$ documents in Table~\ref{table.top80} and top-$100$ documents in Table~\ref{table.top100}.

\setlength{\tabcolsep}{4pt}
\begin{table*}[h]
\caption{nDCG@10 on BEIR benchmark top-$60$.}
\label{table.top60}
\begin{center}
\scriptsize
\begin{tabular}{ | c | c | c c c | c c c | c c c | c c c | }
\hline
Task & Dense & \multicolumn{3}{c|}{Mistral 7B} & \multicolumn{3}{c|}{Llama-3.1 8B} & \multicolumn{3}{c|}{Qwen-3 4B} & \multicolumn{3}{c|}{Granite-3.2 8B} \\\cline{3-14}
& baseline & ICR & QR-R & CoRe-R & ICR & QR-R & CoRe-R & ICR & QR-R & CoRe-R & ICR & QR-R & CoRe-R \\
\hline\hline
TREC-COVID & 63.1 & 71.5 & 74.3 & \textbf{74.6} & 76.9 & 77.5 & \textbf{77.9} & 76.2 & 76.8 & \textbf{77.7} & \textbf{73.7} & 71.7 & 71.9 \\
NFCorpus & 33.7 & 32.7 & 34.3 & \textbf{34.8} & 34.5 & 35.5 & \textbf{36.8} & 35.7 & 35.6 & \textbf{36.7} & 32.4 & \textbf{33.0} & \textbf{33.0} \\
DBPedia & 36.0 & 35.3 & 36.9 & \textbf{37.5} & 38.6 & 39.3 & \textbf{39.8} & 38.9 & 39.1 & \textbf{40.1} & 35.9 & 35.3 & \textbf{36.6} \\
SciFact & 71.3 & 72.1 & 71.3 & \textbf{73.6} & 74.3 & 74.5 & \textbf{75.3} & 73.4 & 74.6 & \textbf{76.1} & 75.1 & 74.2 & \textbf{74.5} \\
SciDocs & 22.5 & 16.7 & 17.1 & \textbf{19.3} & 18.5 & 18.6 & \textbf{19.6} & 21.0 & 19.8 & \textbf{21.6} & 19.4 & 19.5 & \textbf{19.6} \\
FiQA & 36.9 & 37.0 & 39.6 & \textbf{41.6} & 42.0 & 42.7 & \textbf{44.2} & 40.6 & 41.3 & \textbf{43.9} & 41.1 & 40.9 & \textbf{41.9} \\
NQ & 51.6 & 53.1 & 54.9 & \textbf{56.8} & 61.0 & \textbf{63.3} & 63.2 & 61.5 & 62.9 & \textbf{64.1} & 59.5 & 57.9 & \textbf{60.5} \\
FEVER & 85.5 & 87.4 & 87.0 & \textbf{88.4} & \textbf{89.1} & 88.2 & 88.1 & \textbf{89.2} & 86.8 & 88.5 & \textbf{88.4} & 88.0 & 86.6 \\
Climate-FEVER & 30.3 & 21.4 & 20.8 & \textbf{23.2} & 22.3 & 21.7 & \textbf{22.6} & 23.1 & 23.5 & \textbf{25.1} & \textbf{21.9} & 21.1 & 21.0 \\
HotpotQA & 62.9 & 72.2 & 72.5 & \textbf{74.2} & 74.9 & 74.6 & \textbf{75.0} & 73.6 & 74.5 & \textbf{75.3} & 72.0 & 73.3 & \textbf{73.6} \\
Touche & 24.0 & 23.5 & 25.8 & \textbf{27.6} & 26.9 & 27.4 & \textbf{28.2} & 23.5 & 22.4 & \textbf{27.2} & 23.5 & 25.2 & \textbf{25.6} \\
MSMARCO & 30.7 & 28.8 & \textbf{31.1} & 31.0 & 32.8 & 34.5 & \textbf{34.8} & 33.2 & 33.5 & \textbf{34.4} & 30.7 & 30.4 & \textbf{31.8} \\
Quora & 86.7 & 81.2 & 71.4 & \textbf{85.1} & 78.0 & 72.9 & \textbf{81.5} & 72.7 & 63.0 & \textbf{81.4} & \textbf{79.5} & 70.1 & 74.5 \\
ArguAna & 56.4 & 46.6 & 51.2 & \textbf{53.3} & 43.0 & 50.5 & \textbf{55.1} & 37.4 & 39.1 & \textbf{52.5} & 54.7 & 54.6 & \textbf{56.7} \\
CQADupstack & 44.3 & 38.3 & 40.3 & \textbf{41.2} & 41.7 & 43.1 & \textbf{43.4} & 42.7 & 42.1 & \textbf{44.9} & 42.3 & 43.0 & \textbf{43.6} \\
\hline
Average & 49.1 & 47.8 & 48.6 & \textbf{50.8} & 50.2 & 50.9 & \textbf{52.4} & 49.5 & 49.0 & \textbf{52.6} & 50.0 & 49.2 & \textbf{50.1} \\
\hline
\end{tabular}
\end{center}
\end{table*}

\setlength{\tabcolsep}{4pt}
\begin{table}[h]
\caption{nDCG@10 on the BEIR benchmark top-$80$.}
\label{table.top80}
\begin{center}
\scriptsize
\begin{tabular}{ | c | c | c c c | c c c | c c c | c c c | }
\hline
Task & Dense & \multicolumn{3}{c|}{Mistral 7B} & \multicolumn{3}{c|}{Llama-3.1 8B} & \multicolumn{3}{c|}{Qwen-3 4B} & \multicolumn{3}{c|}{Granite-3.2 8B} \\\cline{3-14}
& baseline & ICR & QR-R & CoRe-R & ICR & QR-R & CoRe-R & ICR & QR-R & CoRe-R & ICR & QR-R & CoRe-R \\
\hline\hline
TREC-COVID & 63.1 & 69.6 & 74.0 & \textbf{74.8} & \textbf{78.4} & 78.3 & 78.2 & 76.8 & \textbf{77.9} & 77.1 & 73.4 & 70.4 & \textbf{73.7} \\
NFCorpus & 33.7 & 32.7 & 33.7 & \textbf{33.9} & 33.4 & 35.3 & \textbf{36.8} & 35.2 & 35.4 & \textbf{37.0} & 31.6 & 31.8 & \textbf{32.6} \\
DBPedia & 36.0 & 34.6 & 36.8 & \textbf{37.6} & 39.0 & 39.5 & \textbf{40.0} & 39.2 & 39.2 & \textbf{39.9} & 36.7 & 35.5 & \textbf{36.4} \\
SciFact & 71.3 & 70.6 & 67.1 & \textbf{71.5} & 73.5 & 74.0 & \textbf{76.0} & 75.0 & 75.2 & \textbf{76.0} & \textbf{74.5} & 73.8 & 74.1 \\
SciDocs & 22.5 & 16.4 & 17.1 & \textbf{19.0} & 18.1 & 18.2 & \textbf{19.4} & 21.0 & 20.0 & \textbf{21.7} & 19.4 & \textbf{19.5} & 19.1 \\
FiQA & 36.9 & 36.1 & 40.1 & \textbf{41.7} & 41.3 & 42.3 & \textbf{43.5} & 40.8 & 42.4 & \textbf{44.4} & 40.4 & 39.8 & \textbf{41.1} \\
NQ & 51.6 & 51.3 & 54.5 & \textbf{56.3} & 60.8 & \textbf{63.5} & 62.8 & 61.5 & 62.1 & \textbf{63.9} & 59.3 & 57.7 & \textbf{60.3} \\
FEVER & 85.5 & 87.3 & 86.9 & \textbf{88.5} & 86.0 & 86.1 & \textbf{87.9} & \textbf{89.3} & 86.0 & 88.6 & \textbf{87.3} & 87.1 & 87.1 \\
Climate-FEVER & 30.3 & 21.1 & 20.9 & \textbf{22.7} & 22.2 & \textbf{22.5} & 22.4 & 25.2 & 25.1 & \textbf{25.6} & \textbf{22.2} & 21.2 & 21.3 \\
HotpotQA & 62.9 & 72.5 & 72.9 & \textbf{74.8} & 75.6 & 75.4 & \textbf{75.8} & 74.1 & 75.2 & \textbf{76.0} & \textbf{75.6} & 74.6 & 74.2 \\
Touche & 24.0 & 22.9 & 26.6 & \textbf{28.5} & 27.9 & 27.1 & \textbf{28.7} & 26.9 & 27.3 & \textbf{27.4} & 21.8 & 23.8 & \textbf{24.2} \\
MSMARCO & 30.7 & 27.9 & 28.7 & \textbf{30.4} & 32.7 & 34.4 & \textbf{34.7} & 33.4 & 34.0 & \textbf{34.4} & 30.7 & 30.2 & \textbf{32.2} \\
Quora & 86.7 & 81.3 & 71.7 & \textbf{85.1} & 79.0 & 72.6 & \textbf{82.4} & 73.9 & 61.5 & \textbf{81.5} & 73.5 & 72.6 & \textbf{74.9} \\
ArguAna & 56.4 & 48.2 & 52.3 & \textbf{53.1} & 43.2 & 50.7 & \textbf{55.2} & 38.6 & 41.1 & \textbf{51.7} & 54.3 & 55.1 & \textbf{55.7} \\
CQADupstack & 44.3 & 38.0 & 40.1 & \textbf{40.7} & 41.6 & 42.9 & \textbf{43.2} & 42.7 & 43.1 & \textbf{45.0} & 42.2 & 42.0 & \textbf{42.9} \\
\hline
Average & 49.1 & 47.4 & 48.2 & \textbf{50.6} & 50.1 & 50.8 & \textbf{52.5} & 50.2 & 49.7 & \textbf{52.7} & 49.5 & 49.0 & \textbf{50.0} \\
\hline
\end{tabular}
\end{center}
\end{table}

\setlength{\tabcolsep}{4pt}
\begin{table}[!h]
\caption{nDCG@10 on the BEIR benchmark top-$100$.}
\label{table.top100}
\begin{center}
\scriptsize
\begin{tabular}{ | c | c | c c c | c c c | c c c | c c c | }
\hline
Task & Dense & \multicolumn{3}{c|}{Mistral 7B} & \multicolumn{3}{c|}{Llama-3.1 8B} & \multicolumn{3}{c|}{Qwen-3 4B} & \multicolumn{3}{c|}{Granite-3.2 8B} \\\cline{3-14}
& baseline & ICR & QR-R & CoRe-R & ICR & QR-R & CoRe-R & ICR & QR-R & CoRe-R & ICR & QR-R & CoRe-R \\
\hline\hline
TREC-COVID & 63.1 & 66.8 & 73.0 & \textbf{73.7} & 76.2 & 79.2 & \textbf{79.6} & 77.2 & 76.4 & \textbf{77.7} & 72.9 & 71.6 & \textbf{75.0} \\
NFCorpus & 33.7 & 31.5 & 31.3 & \textbf{32.5} & 33.2 & 35.2 & \textbf{36.8} & 35.6 & 34.3 & \textbf{36.9} & 31.9 & 31.8 & \textbf{32.7} \\
DBPedia & 36.0 & 33.6 & 36.7 & \textbf{37.5} & 38.9 & 39.7 & \textbf{39.8} & 39.1 & 39.0 & \textbf{40.3} & 36.5 & 36.1 & \textbf{37.4} \\
SciFact & 71.3 & 62.9 & 58.2 & \textbf{66.0} & 74.1 & 74.8 & \textbf{76.7} & 75.8 & 75.7 & \textbf{76.0} & \textbf{74.0} & \textbf{74.0} & 73.8 \\
SciDocs & 22.5 & 16.5 & 16.5 & \textbf{18.6} & 17.5 & 18.2 & \textbf{19.3} & 20.9 & 19.3 & \textbf{21.5} & 19.4 & \textbf{19.5} & 19.3 \\
FiQA & 36.9 & 35.6 & 39.5 & \textbf{40.3} & 41.0 & 42.0 & \textbf{43.9} & 41.2 & 42.3 & \textbf{44.3} & 40.4 & 39.7 & \textbf{41.0} \\
NQ & 51.6 & 50.2 & 53.8 & \textbf{55.5} & 60.7 & 60.9 & \textbf{62.4} & 61.4 & 62.4 & \textbf{64.0} & 58.8 & 57.0 & \textbf{60.3} \\
FEVER & 85.5 & 87.1 & 86.4 & \textbf{88.2} & \textbf{88.9} & 88.0 & 87.8 & \textbf{89.3} & 86.2 & 88.7 & 86.5 & 87.1 & \textbf{87.4} \\
Climate-FEVER & 30.3 & 21.4 & 21.5 & \textbf{23.1} & 22.3 & 22.1 & \textbf{22.6} & 25.4 & 25.0 & \textbf{26.1} & \textbf{22.1} & 21.0 & 21.9 \\
HotpotQA & 62.9 & 72.7 & 73.2 & \textbf{75.1} & 76.0 & 75.8 & \textbf{76.1} & 74.6 & 75.6 & \textbf{76.7} & 73.4 & 74.1 & \textbf{75.2} \\
Touche & 24.0 & 21.6 & 26.4 & \textbf{27.1} & 26.9 & 26.1 & \textbf{27.3} & 25.2 & 26.1 & \textbf{27.6} & 21.2 & 23.1 & \textbf{23.7} \\
MSMARCO & 30.7 & 27.3 & 28.5 & \textbf{30.2} & 32.3 & \textbf{34.2} & \textbf{34.2} & 33.0 & 34.0 & \textbf{34.4} & 30.2 & 31.0 & \textbf{32.1} \\
Quora & 86.7 & 81.5 & 72.4 & \textbf{85.1} & 78.9 & 71.7 & \textbf{82.0} & 74.4 & 64.0 & \textbf{81.5} & 73.7 & 74.9 & \textbf{75.7} \\
ArguAna & 56.4 & 48.5 & \textbf{52.4} & \textbf{52.4} & 42.5 & 50.3 & \textbf{54.9} & 37.1 & 38.2 & \textbf{51.0} & 53.9 & 54.4 & \textbf{55.2} \\
CQADupstack & 44.3 & 37.6 & 39.2 & \textbf{39.8} & 41.3 & 42.5 & \textbf{43.0} & 42.7 & 43.8 & \textbf{45.1} & 41.9 & 42.0 & \textbf{42.7} \\
\hline
Average & 49.1 & 46.3 & 47.3 & \textbf{49.7} & 50.0 & 50.7 & \textbf{52.4} & 50.2 & 49.5 & \textbf{52.8} & 49.1 & 49.1 & \textbf{50.2} \\
\hline
\end{tabular}
\end{center}
\end{table}

\setlength{\tabcolsep}{4pt}
\begin{table}[!h]
\caption{nDCG@10 on the BEIR benchmark for Phi-4 across different top-$k$.}
\label{table.phi4}
\begin{center}
\scriptsize
\begin{tabular}{ | c | c | c c c | c c c | c c c | c c c | }
\hline
Task & Dense & \multicolumn{3}{c|}{top-40} & \multicolumn{3}{c|}{top-60} & \multicolumn{3}{c|}{top-80} & \multicolumn{3}{c|}{top-100} \\\cline{3-14}
& baseline & ICR & QR-R & CoRe-R & ICR & QR-R & CoRe-R & ICR & QR-R & CoRe-R & ICR & QR-R & CoRe-R \\
\hline\hline
TREC-COVID    & 63.1 & 75.2 & \textbf{76.8} & 76.0 & 73.8 & \textbf{77.0} & \textbf{77.0} & 73.1 & \textbf{77.5} & \textbf{77.5} & 71.0 & 79.0 & \textbf{79.2} \\
NFCorpus      & 33.7 & 31.6 & 35.5 & \textbf{35.6} & 27.2 & 34.6 & \textbf{34.8} & 23.8 & \textbf{32.9} & 32.6 & 18.3 & 23.3 & \textbf{23.4} \\
DBPedia       & 36.0 & 36.9 & 39.4 & \textbf{39.6} & 36.4 & 40.0 & \textbf{40.3} & 35.8 & \textbf{40.5} & \textbf{40.5} & 35.3 & \textbf{40.5} & \textbf{40.5} \\
SciFact       & 71.3 & 71.7 & 74.2 & \textbf{74.7} & 65.2 & 70.3 & \textbf{70.9} & 53.6 & 60.2 & \textbf{61.6} & 41.3 & 45.4 & \textbf{45.9} \\
SciDocs       & 22.5 & 17.8 & 21.0 & \textbf{21.1} & 17.5 & \textbf{20.7} & \textbf{20.7} & 15.8 & 19.0 & \textbf{19.8} & 10.1 & 15.1 & \textbf{15.2} \\
FiQA          & 36.9 & 37.2 & 43.8 & \textbf{44.1} & 36.0 & 42.8 & \textbf{43.0} & 33.3 & \textbf{42.0} & 41.8 & 27.7 & 36.3 & \textbf{36.8} \\
NQ            & 51.6 & 57.4 & 63.3 & \textbf{63.6} & 56.9 & 63.1 & \textbf{63.7} & 56.1 & 63.2 & \textbf{63.3} & 54.8 & 62.8 & \textbf{62.9} \\
FEVER         & 85.5 & 89.1 & \textbf{89.6} & \textbf{89.6} & 89.0 & \textbf{89.6} & \textbf{89.6} & 88.5 & 89.2 & \textbf{89.3} & 87.2 & \textbf{88.8} & 88.6 \\
Climate-FEVER & 30.3 & 22.2 & \textbf{24.5} & 24.3 & 22.4 & \textbf{24.9} & 24.5 & 21.5 & 24.0 & \textbf{24.3} & 17.9 & 21.3 & \textbf{21.7} \\
HotpotQA      & 62.9 & 73.3 & 74.2 & \textbf{74.5} & 74.4 & 75.6 & \textbf{75.7} & 74.9 & \textbf{76.2} & \textbf{76.2} & 75.4 & 76.7 & \textbf{76.8} \\
Touche        & 24.0 & 21.8 & 25.4 & \textbf{25.7} & 19.2 & 25.3 & \textbf{25.6} & 19.3 & 26.3 & \textbf{26.6} & 14.7 & \textbf{24.6} & 24.3 \\
MSMARCO       & 30.7 & 30.6 & 34.6 & \textbf{34.8} & 30.1 & 34.5 & \textbf{34.7} & 29.7 & \textbf{34.6} & \textbf{34.6} & 28.9 & 34.4 & \textbf{34.5} \\
Quora         & 86.7 & 74.2 & 76.5 & \textbf{80.2} & 75.2 & 70.8 & \textbf{80.5} & 74.6 & 74.4 & \textbf{78.6} & 75.3 & 76.1 & \textbf{78.4} \\
ArguAna       & 56.4 & 41.7 & 52.2 & \textbf{52.7} & 40.7 & 49.8 & \textbf{51.7} & 36.6 & 47.8 & \textbf{48.2} & 17.1 & 32.2 & \textbf{32.3} \\
CQADupstack   & 44.3 & 39.4 & 44.0 & \textbf{44.3} & 37.3 & 42.9 & \textbf{43.5} & 36.3 & 42.1 & \textbf{42.4} & 32.9 & \textbf{40.9} & 39.6 \\
\hline
Average       & 49.1 & 48.0 & 51.7 & \textbf{52.1} & 46.7 & 50.8 & \textbf{51.7} & 44.9 & 49.9 & \textbf{50.5} & 40.5 & 46.4 & \textbf{46.7} \\
\hline
\end{tabular}
\end{center}
\end{table}

In all context length settings, CoRe-R outperforms the baselines across all models. Similar to the reranking results for top-$40$ in the main text, CoRe-R shows prominent improvement in all datasets for Mistral 7B. For Llama-3.1 8B, CoRe-R even shows further improvement with longer context length on many datasets compared to top-$40$, while both baselines ICR and QR-R show accuracy degradation. These results widen the gap between CoRe-R and the baselines, demonstrating the superior robustness of CoRe-R in long context settings.

Table~\ref{table.phi4} presents the reranking results using \textit{microsoft/phi-4} \citep{abdin2024phi} across different top-$k$. We found that Phi shows strong performance for smaller top-$k$ but tends to degrade as $k$ increases. There is a noticeable degradation in top-100 where the reranking accuracy significantly underperform the retriever baseline. This degradation stems from Phi's limited context length support (16k), which is much smaller compared to Mistral at 32k, Llama-3.1 at 128k, and Qwen at 262k. As more recent LLMs increasingly support extended contexts, this context length limitation will no longer be an issue.

\subsection{Results on multi-hop tasks}
\label{app.multihop}

We conduct additional experiment on two multi-hop tasks MuSiQue \citep{trivedi2022musique} and CLIPPER \citep{pham2025clipper}. We follow the same setup in \citep{chenattention} for MuSiQue dataset and \citep{zhang2025query} for CLIPPER dataset. Table~\ref{table.multihop} reports Recall@2 and Recall@5 for both multi-hop datasets with four language models. Overall, CoRe-R stands out with the best reranking accuracy in both tasks across all models with only one exception of CLIPPER Recall@5 in Llama-3.1 8B.

\setlength{\tabcolsep}{4pt}
\begin{table}[h]
\caption{Reranking performance on multi-hop datasets (top-$40$).}
\label{table.multihop}
\begin{center}
\scriptsize
\begin{tabular}{ | c | c | c c c | c c c | c c c | c c c | }
\hline
Task & Dense & \multicolumn{3}{c|}{Mistral 7B} & \multicolumn{3}{c|}{Llama-3.1 8B} & \multicolumn{3}{c|}{Qwen-4 3B} & \multicolumn{3}{c|}{Granite-3.2 8B} \\\cline{3-14}
& baseline & ICR & QR-R & CoRe-R & ICR & QR-R & CoRe-R & ICR & QR-R & CoRe-R & ICR & QR-R & CoRe-R \\
\hline\hline
MuSiQue (R@2) & 37.9        & 40.0 & 41.9 & \textbf{43.9}         & 44.5 & 45.9 & \textbf{46.7}        & 45.6 & 46.0 & \textbf{47.3}    & 45.7 & 46.1 & \textbf{46.8} \\
MuSiQue (R@5) & 49.2        & 53.9 & 54.4 & \textbf{56.2}         & 57.0 & 58.3 & \textbf{58.9}        & 58.8 & 59.2 & \textbf{60.8}    & 56.4 & 57.2 & \textbf{58.4} \\
CLIPPER (R@2) & 5.1         & 24.2 & 25.9 & \textbf{26.5}         & 27.8 & 29.0 & \textbf{29.8}        & 28.3 & 28.3 & \textbf{29.5}    & 26.6 & 26.7 & \textbf{27.5} \\
CLIPPER (R@5) & 22.0        & 43.0 & 45.1 & \textbf{45.7}         & 45.7 & \textbf{48.8} & 48.6        & 46.7 & 47.2 & \textbf{48.0}    & 46.1 & 44.6 & \textbf{47.1} \\
\hline
\end{tabular}
\end{center}
\end{table}

\subsection{Results on length bias study}
\label{app.length.bias}

To understand the presence of length bias issue, we sample (query, documents) pairs from all BEIR datasets that satisfy the following extreme condition: the ratio of the gold document length and the longest irrelevant document is at most 0.2, which totals to over $4000$ samples. We run reranking tests using Llama-3.1 8B with and without contextual calibration, and we include the generative baseline RankGPT (GPT-4) for comparison. We report the win rate for which the short gold document scores higher than the longest irrelevant document, and the overall nDCG@10 scores in Table~\ref{table.length.bias.full}.

\begin{table}[h]
\caption{Win rate and nDCG@10 on extreme cases in BEIR. ICR, QR-R and CoRe-R use Llama.}
\label{table.length.bias.full}
\small
\centering
\begin{tabular}{| c | c c c c |}
    \hline
    Reranker & Win rate & Win rate & nDCG@10 & nDCG@10 \\
    & (w/o calib) & (w/ calib) & (w/o calib) & (w/ calib) \\
    \hline\hline
    RankGPT-4 & 67.05 & N/A & 30.10 & N/A \\
    ICR & 31.25 & 59.09 & 10.81 & 25.00 \\
    QR-R & 57.39 & 67.89 & 25.67 & 28.75 \\
    CoRe-R & 57.89 & 69.32 & 26.92 & 29.02 \\
    \hline
\end{tabular}
\end{table}

There are two key observations here. First, ICR without calibration severely suffers from length bias issue, and QR-R and CoRe-R are generally more robust. This makes intuitive sense because ICR applies a noisy aggregation over all heads, while QR-R and CoRe-R applies a selective aggregation learned from the head detection step. This result may suggest that the head detection process can act as a form of calibration by keeping strong attention heads and remove heads that are noisy and biased. Second, CoRe-R achieves the highest win rate when calibration is applied, even surpassing RankGPT, indicating strong robustness in this deliberately adversarial short-gold/long-negative setting. It provides direct evidence that CoRe-R is not systematically biased toward longer passages.


\subsection{Sensitivity to the number of CoRe heads}
\label{app.optimal.core.head}
\begin{figure*}[h]
\centering
\begin{subfigure}{0.329\textwidth}
\centering
  \includegraphics[width=\textwidth]{figures/mistral_nq_allhead.pdf}
  \caption{Mistral 7B on NQ}
\end{subfigure}
\begin{subfigure}{0.329\textwidth}
\centering
  \includegraphics[width=\textwidth]{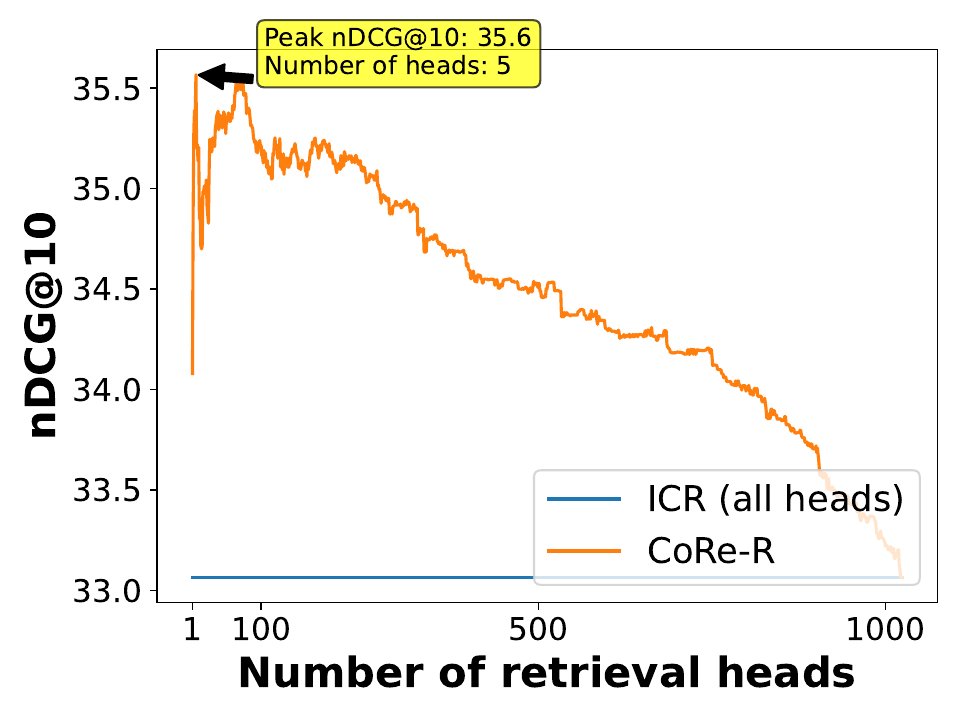}
  \caption{Mistral 7B on NFCorpus}
\end{subfigure}
\begin{subfigure}{0.329\textwidth}
\centering
  \includegraphics[width=\textwidth]{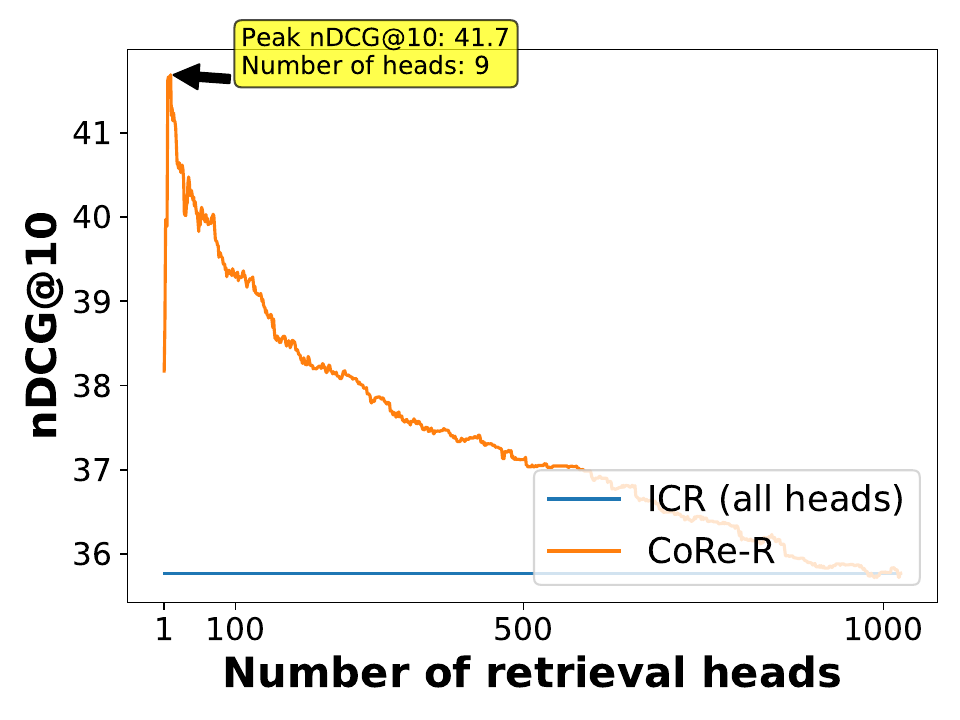}
  \caption{Mistral 7B on FiQA}
\end{subfigure}

\begin{subfigure}{0.329\textwidth}
\centering
  \includegraphics[width=\textwidth]{figures/llama_nq_allhead.pdf}
  \caption{Llama-3.1 8B on NQ}
\end{subfigure}
\begin{subfigure}{0.329\textwidth}
\centering
  \includegraphics[width=\textwidth]{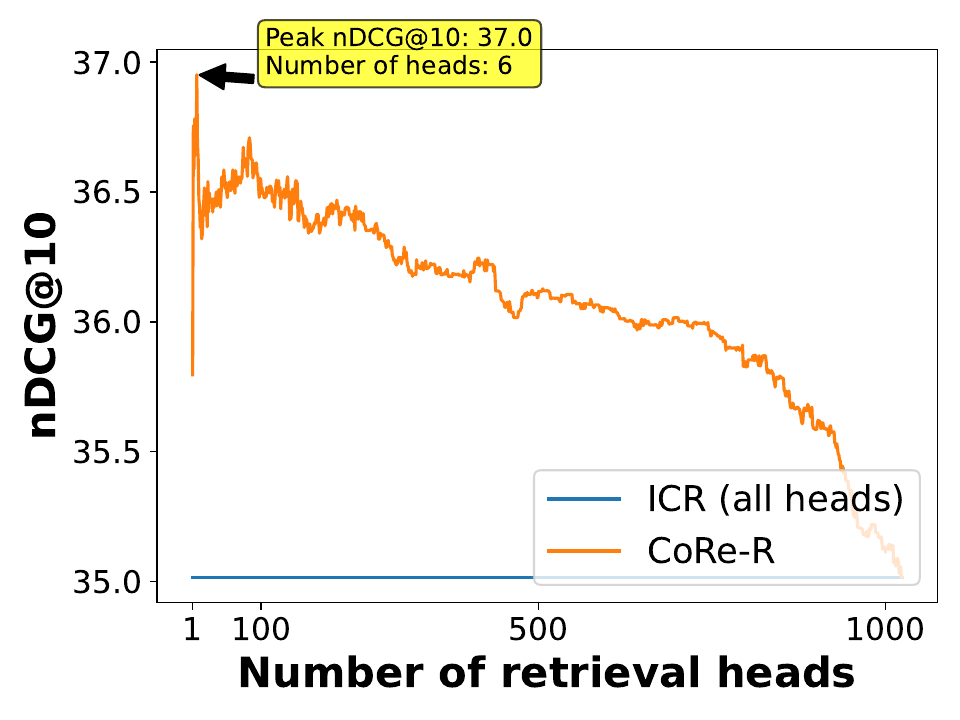}
  \caption{Llama-3.1 8B on NFCorpus}
\end{subfigure}
\begin{subfigure}{0.329\textwidth}
\centering
  \includegraphics[width=\textwidth]{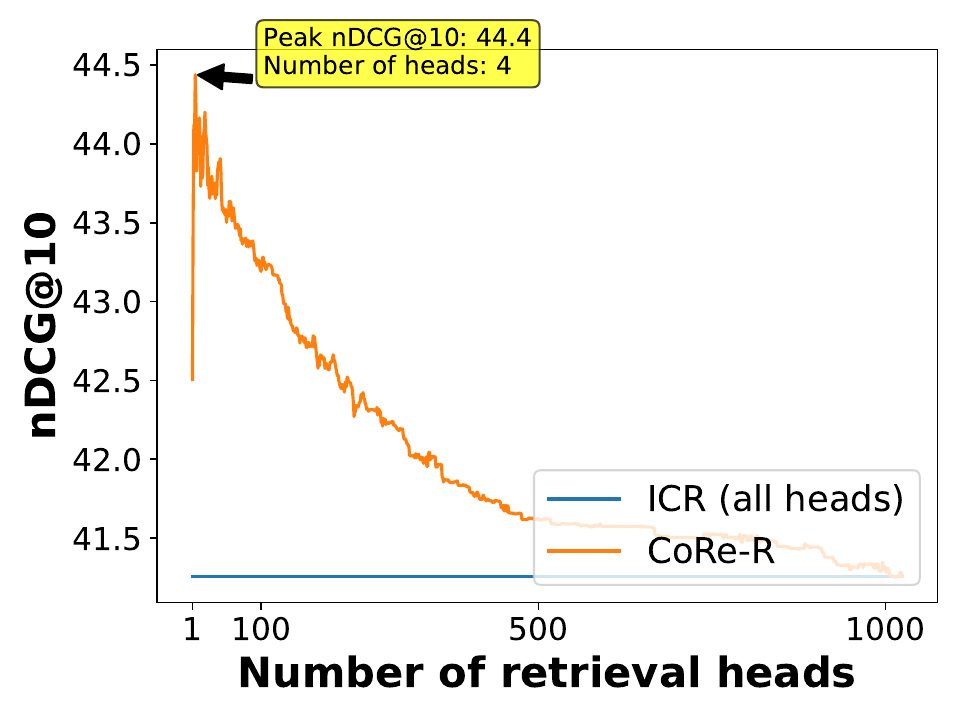}
  \caption{Llama-3.1 8B on FiQA}
\end{subfigure}
\caption{nDCG@10 with CoRe-R across all number of CoRe heads.}
\label{app.fig.allhead}
\end{figure*}

We study the effect of varying the number of CoRe heads across three datasets (NQ, NFCorpus, and FiQA) using Mistral 7B and Llama-3.1 8B. Figure~\ref{app.fig.allhead} reports reranking performance as additional CoRe heads are incrementally aggregated.

Across all datasets and models, reranking accuracy consistently peaks using only a small number of top-ranked CoRe heads, typically within the top-$10$. Adding lower-ranked heads gradually degrades performance, suggesting that retrieval-relevant signals are highly concentrated while weaker heads contribute increasingly noisy or redundant information. The consistency of this trend across diverse retrieval settings further supports the robustness of CoRe head detection and indicates that only a small fixed number of CoRe heads is sufficient in practice.

\subsection{Sensitivity to the detection dataset}
\label{app.msmarco}

To evaluate the sensitivity of CoRe-R to the detection dataset, we repeat the head detection procedure using a subset of MSMARCO instead of NQ. The resulting CoRe heads are highly consistent with those detected from NQ, with $7$ out of the top-$8$ heads overlapping between the two settings.

Table~\ref{table.hard.negative.data} reports the reranking performance using the MSMARCO-detected CoRe heads. The results remain highly consistent across datasets and models, suggesting that CoRe head detection is robust and does not strongly depend on the specific detection dataset used.

\begin{table}[h]
\caption{nDCG@10 of CoRe-R on BEIR benchmark with different hard negatives data.}
\label{table.hard.negative.data}
\begin{center}
\small
\begin{tabular}{ | c | c | c c | c c | }
\hline
Task & Dense & \multicolumn{2}{c|}{Mistral 7B} & \multicolumn{2}{c|}{Llama-3.1 8B} \\\cline{3-6}
& baseline & NQ & MSMARCO & NQ & MSMARCO \\
\hline\hline
TREC-COVID & 63.1 & 73.8 & 74.5 & 75.7 & 75.6 \\
NFCorpus & 33.7 & 35.2 & 35.6 & 36.7 & 36.5 \\
DBPedia & 36.0 & 37.5 & 38.1 & 39.4 & 39.4 \\
SciFact & 71.3 & 74.7 & 73.6 & 75.1 & 75.8 \\
SciDocs & 22.5 & 19.5 & 19.0 & 20.2 & 20.0 \\
FiQA & 36.9 & 41.4 & 41.3 & 44.1 & 43.9 \\
NQ & 51.6 & 58.1 & 58.3 & 63.2 & 63.3 \\
FEVER & 85.5 & 88.7 & 88.3 & 88.4 & 88.7 \\
Climate-FEVER & 30.3 & 23.3 & 22.7 & 22.7 & 22.7 \\
HotpotQA & 62.9 & 73.4 & 73.0 & 73.8 & 73.9 \\
Touche & 24.0 & 26.2 & 27.4 & 26.4 & 26.3 \\
MSMARCO & 30.7 & 31.7 & 32.3 & 34.9 & 35.1 \\
Quora & 86.7 & 85.2 & 83.8 & 82.3 & 82.0 \\
ArguAna & 56.4 & 52.7 & 52.2 & 55.0 & 54.7 \\
CQADupstack & 44.3 & 41.6 & 41.4 & 43.9 & 43.9 \\
\hline
Average & 49.1 & 50.9 & 50.8 & 52.1 & 52.1 \\
\hline
\end{tabular}
\end{center}
\end{table}


\subsection{Sensitivity to hard-negative configurations}
\label{app.negative.exp}

We conduct additional experiments using different hard-negative constructions, including different retrievers (BM25 vs.\ dense retrieval) and different positions of the positive document within the candidate set (early, middle, late, and random). We use seed $42$ for random setting.

\begin{table}[h]
\caption{nDCG@10 of CoRe-R using Llama with different hard-negative configurations (retriever type and positive position). CoRe-R (dense) and CoRe-R (early) use the same CoRe heads used in the main experiment (Table~\ref{table.beir}).}
\label{table.hard.negatives}
\begin{center}
\small
\begin{tabular}{ | c | c c | c c c c | }
\hline
Task & \multicolumn{2}{c|}{Retriever} & \multicolumn{4}{c|}{Positive positions} \\\cline{2-7}
& dense & BM25 & early & middle & late & random \\
\hline\hline
TREC-COVID & 75.7 & 76.2 & 75.7 & 76.0 & 75.9 & 76.0 \\
NFCorpus & 36.7 & 36.5 & 36.7 & 36.4 & 36.5 & 36.4 \\
DBPedia & 39.4 & 39.0 & 39.4 & 39.5 & 39.5 & 39.5 \\
SciFact & 75.1 & 75.2 & 75.1 & 75.0 & 75.3 & 75.0 \\
SciDocs & 20.2 & 19.9 & 20.2 & 19.8 & 19.5 & 19.8 \\
FiQA & 44.1 & 44.0 & 44.1 & 44.0 & 44.2 & 44.0 \\
NQ & 63.2 & 62.9 & 63.2 & 63.3 & 63.0 & 63.3 \\
FEVER & 88.4 & 88.8 & 88.4 & 88.4 & 88.2 & 88.4 \\
Climate-FEVER & 22.7 & 22.4 & 22.7 & 22.6 & 22.6 & 22.6 \\
HotpotQA & 73.8 & 73.2 & 73.8 & 74.1 & 74.2 & 74.1 \\
Touche & 26.4 & 26.5 & 26.4 & 26.1 & 26.5 & 26.1 \\
MSMARCO & 34.9 & 34.6 & 34.9 & 35.0 & 34.8 & 35.0 \\
Quora & 82.3 & 81.8 & 82.3 & 82.2 & 82.2 & 82.2 \\
ArguAna & 55.0 & 54.7 & 55.0 & 54.8 & 54.9 & 54.8 \\
CQADupstack & 43.9 & 43.9 & 43.9 & 43.9 & 43.9 & 43.9 \\
\hline
Average & 52.1 & 52.0 & 52.1 & 52.0 & 52.1 & 52.0 \\
\hline
\end{tabular}
\end{center}
\end{table}

Across all settings, the detected CoRe heads remain highly consistent. For Llama, the top-$8$ CoRe heads detected under different hard-negative retrievers share $6$ overlapping heads, while the remaining heads still receive high CoRe scores and remain within the top-$16$ ranked heads. Similar behavior is observed for Mistral 7B and Llama-3.1 8B.

We observe the same stability under different positive-document positions. The top-$8$ CoRe heads detected from middle and random positions are identical, while the early and late configurations differ by only a single head. Moreover, the few alternating heads consistently correspond to neighboring heads in layer 14 and remain within the top-$10$ across all configurations.

Table~\ref{table.hard.negatives} reports the corresponding reranking performance. Results remain nearly identical across all settings, indicating that CoRe head detection is robust to both hard-negative construction and positional biases in the detection dataset.

\subsection{Effect of contrastive temperature}
\label{app.temperature}

\begin{figure*}[b]
\centering
\begin{subfigure}{0.329\textwidth}
\centering
  \includegraphics[width=\textwidth]{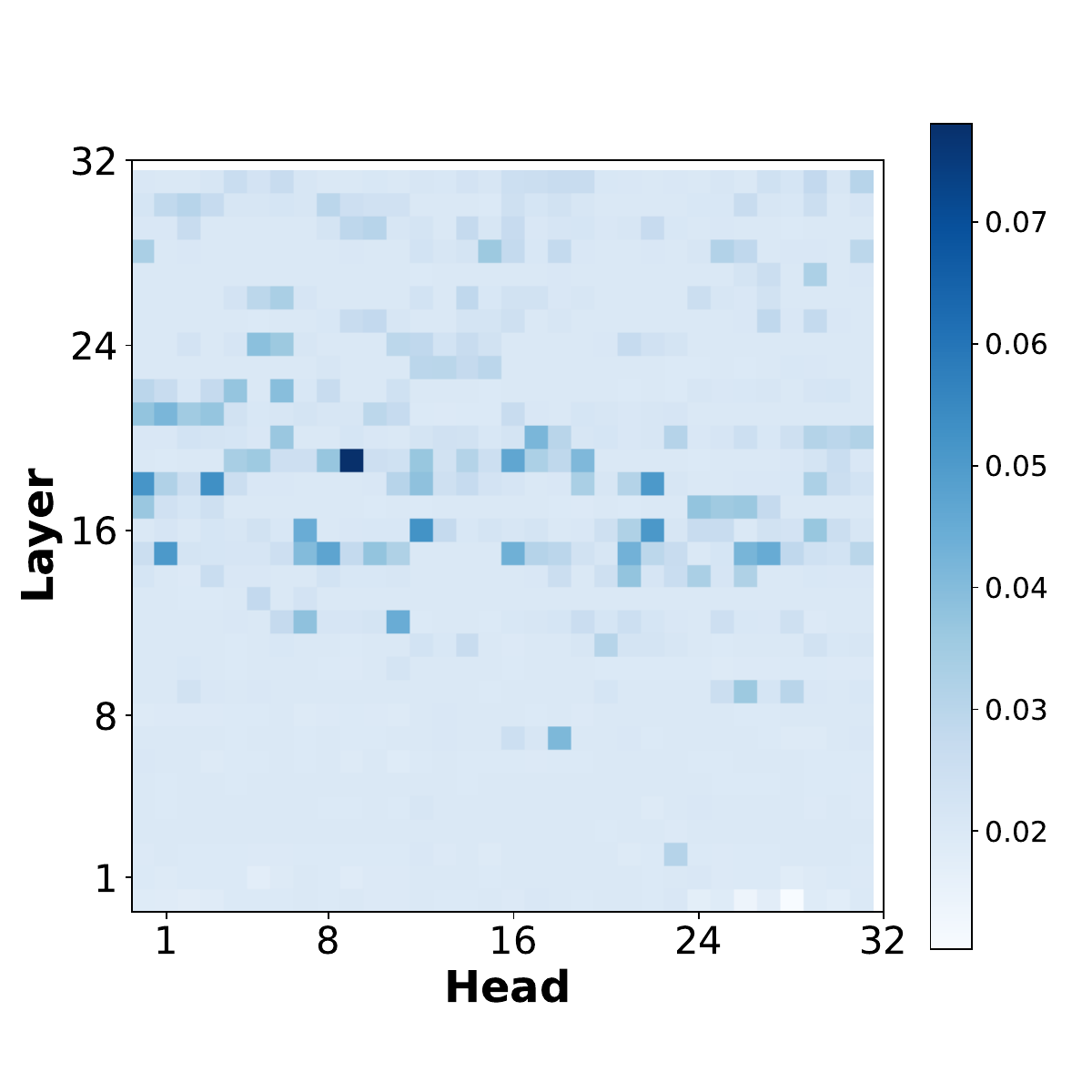}
  \caption{$t = 0.1$}
  \label{app.fig.heatmap.0.1}
\end{subfigure}
\begin{subfigure}{0.329\textwidth}
\centering
  \includegraphics[width=\textwidth]{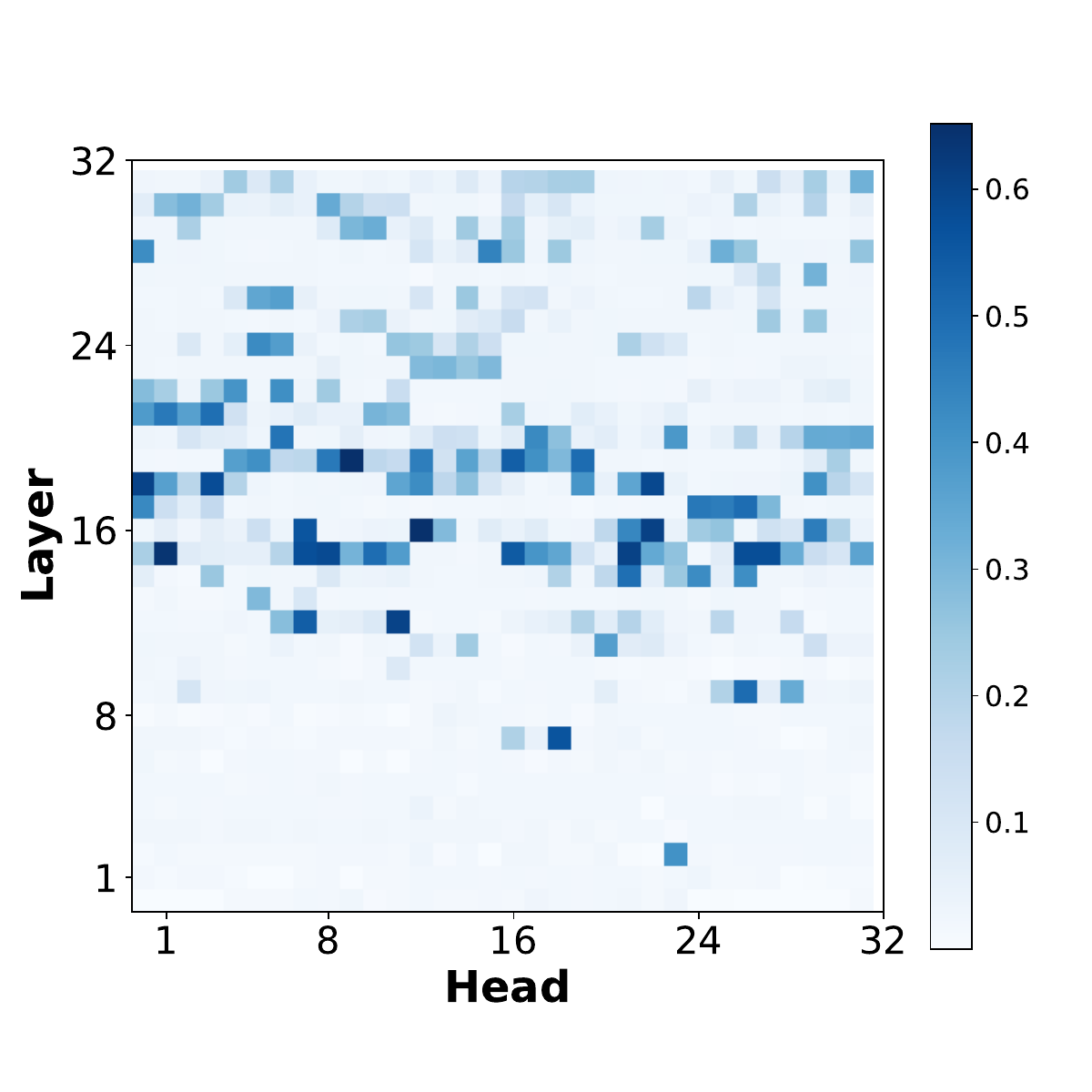}
  \caption{$t=0.01$}
  \label{app.fig.heatmap.0.01}
\end{subfigure}
\begin{subfigure}{0.329\textwidth}
\centering
  \includegraphics[width=\textwidth]{figures/mistral_heatmap.pdf}
  \caption{$t=0.001$}
  \label{app.fig.heatmap.0.001}
\end{subfigure}
\caption{Distribution of $S_{CoRe}$ for all heads in Mistral 7B with different temperature $t$. All temperatures result in similar distribution of CoRe heads with the top few heads lie in the middle layers.}
\label{app.fig.heatmap}
\end{figure*}

We study how different temperature values $t=0.1, 0.01, 0.001$ affect the detected CoRe heads.
While CoRe heads in Llama are consistent across all temperatures, CoRe heads in Mistral highly depend on the temperature (see Figure~\ref{app.fig.heatmap}). We observe that many heads in Mistral attend strongly to both the positive document and hard negatives, thus, higher temperatures $t=0.1, 0.01$ were not able to capture high-quality heads in Mistral.

As a result, the cross-temperature reranking score are nearly identical for LLama, but different for Mistral (see Table~\ref{table.temperature}). While a small amount of temperature tuning is necessary for certain models, the entire tuning process incurs minimal overhead as head detection is a relatively fast one-time process per language model, requiring less than one hour on an H100 (96GB) GPU. Once selected, the same CoRe heads generalize reliably across out-of-domain tasks, multi-hop retrieval, and long-context settings. This demonstrates that CoRe’s performance is not sensitive to temperature in practice, and the detected heads remain robust across applications.

\setlength{\tabcolsep}{4pt}
\begin{table}[h]
\caption{nDCG@10 of CoRe-R on BEIR benchmark top-$40$ with different temperature values.}
\label{table.temperature}
\begin{center}
\scriptsize
\begin{tabular}{ | c | c | c c c | c c c | }
\hline
Task & Dense & \multicolumn{3}{c|}{Mistral 7B} & \multicolumn{3}{c|}{Llama-3.1 8B} \\\cline{3-8}
& baseline & $0.1$ & $0.01$ & $0.001$ & $0.1$ & $0.01$ & $0.001$ \\
\hline\hline
TREC-COVID & 63.1 & 73.6 & 73.8 & 73.8 & 75.7 & 76.3 & 76.3 \\
NFCorpus & 33.7 & 34.9 & 35.6 & 35.2 & 36.7 & 36.6 & 36.6 \\
DBPedia & 36.0 & 37.1 & 37.8 & 37.5 & 39.4 & 39.1 & 39.1 \\
SciFact & 71.3 & 72.2 & 73.0 & 74.7 & 75.1 & 75.9 & 75.9 \\
SciDocs & 22.5 & 17.1 & 18.4 & 19.5 & 20.2 & 20.0 & 20.0 \\
FiQA & 36.9 & 39.5 & 41.2 & 41.4 & 44.1 & 44.3 & 44.3 \\
NQ & 51.6 & 55.6 & 57.8 & 58.1 & 63.2 & 63.6 & 63.6 \\
FEVER & 85.5 & 86.9 & 87.7 & 88.7 & 88.4 & 88.2 & 88.2 \\
Climate-FEVER & 30.3 & 21.1 & 21.9 & 23.3 & 22.7 & 22.9 & 22.9 \\
HotpotQA & 62.9 & 71.6 & 72.4 & 73.4 & 73.8 & 73.4 & 73.4 \\
Touche & 24.0 & 26.6 & 27.1 & 26.2 & 26.4 & 26.9 & 26.9 \\
MSMARCO & 30.7 & 31.7 & 32.3 & 31.7 & 34.9 & 34.8 & 34.8 \\
Quora & 86.7 & 73.4 & 81.5 & 85.2 & 82.3 & 74.5 & 74.5 \\
ArguAna & 56.4 & 50.8 & 52.0 & 52.7 & 55.0 & 54.5 & 54.5 \\
CQADupstack & 44.3 & 40.3 & 41.4 & 41.6 & 43.9 & 43.6 & 43.6 \\
\hline
Average & 49.1 & 48.8 & 50.3 & 50.9 & 52.1 & 51.6 & 51.6 \\
\hline
\end{tabular}
\end{center}
\end{table}

\subsection{Results on layer pruning with Qwen}
\label{app.additional.pruning}

\begin{figure}[h]
    \centering
  \includegraphics[width=0.4\textwidth]{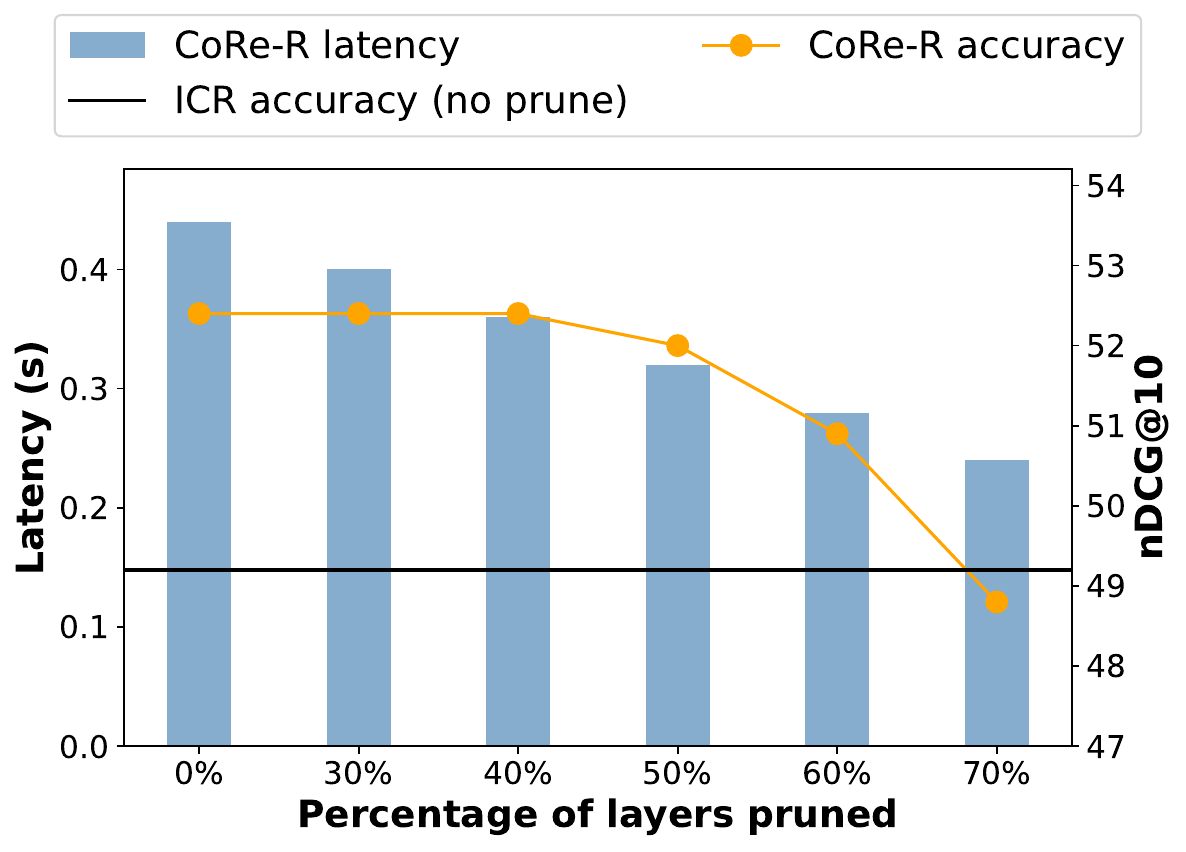}
  \caption{Accuracy vs. latency using Qwen.}
  \label{app.runtime.qwen}
\end{figure}

We also report additional layer pruning experiment for Qwen in Figure~\ref{app.runtime.qwen}. Similar to Mistral and Llama, CoRe-R shows incredible tradeoff in accuracy and efficiency in Qwen. By pruning up to $50\%$ of late layers, CoRe-R maintains near identical accuracy while saving $40\%$ of latency.

\subsection{Layer pruning with NIAH and QR heads}
\label{app.baseline.prune}

\begin{figure*}[h]
\centering
\begin{subfigure}{0.47\textwidth}
\centering
  \includegraphics[width=0.6\textwidth]{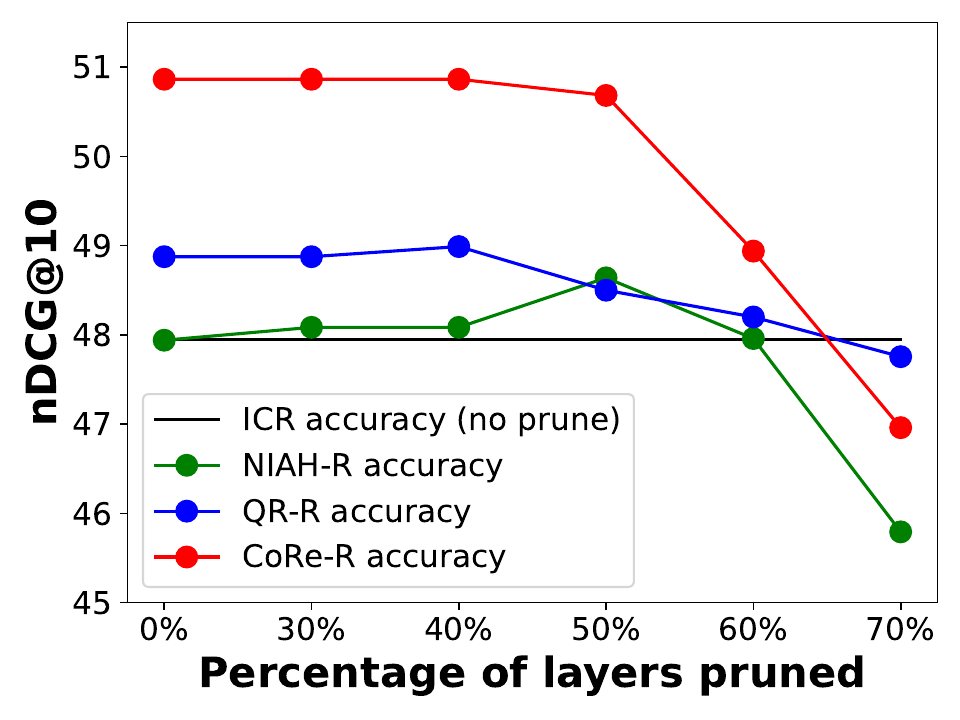}
  \caption{Mistral 7B}
  \label{fig.baseline.prune.mistral}
\end{subfigure}
\begin{subfigure}{0.47\textwidth}
\centering
  \includegraphics[width=0.6\textwidth]{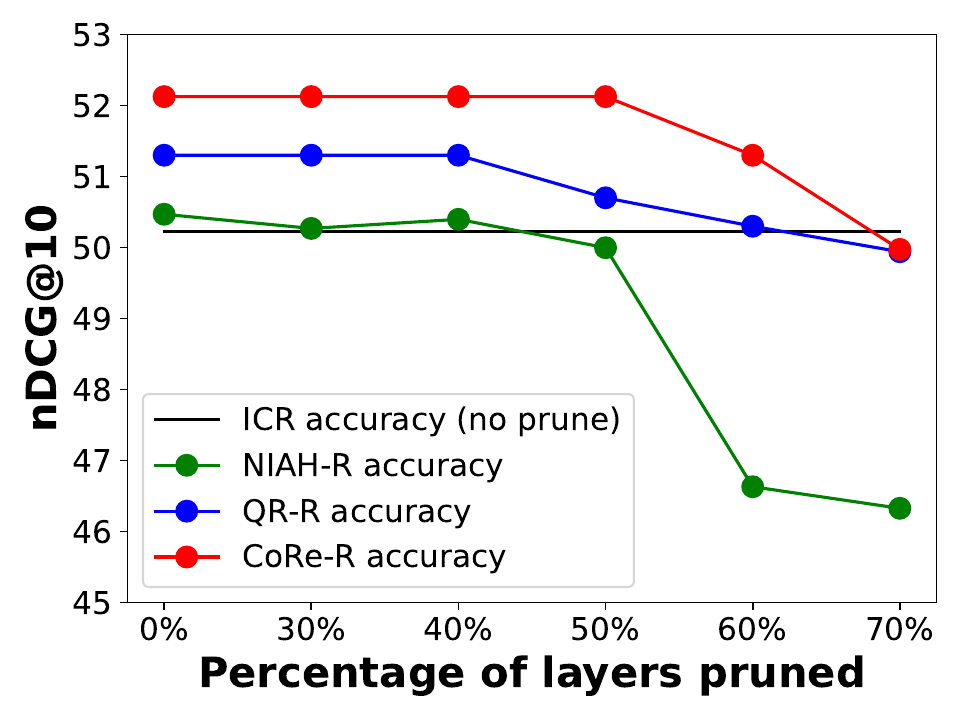}
  \caption{Llama-3.1 8B}
  \label{fig.baseline.prune.llama}
\end{subfigure}
\caption{Average reranking accuracy on BEIR benchmark with different levels of layer pruning.}
\label{fig.baseline.prune}
\end{figure*}

Figure~\ref{fig.baseline.prune} reports the layer pruning results for the baseline rerankers NIAH-R and QR-R. Since both NIAH and QR heads also exhibit partial concentration in middle transformer layers, these methods likewise benefit from moderate layer pruning, achieving improved efficiency under $30$--$40\%$ pruning.

However, both baselines begin to degrade substantially beyond this point, whereas CoRe-R maintains near-identical reranking performance up to $50\%$ pruning. This difference is consistent with the structural localization of the detected heads: while CoRe heads are concentrated almost entirely in early and middle layers, NIAH and QR include a larger number of late-layer heads that become unavailable under aggressive pruning.

Overall, CoRe-R not only achieves stronger reranking accuracy, but also exhibits the best accuracy--efficiency tradeoff due to its highly localized retrieval-specialized head structure.

\subsection{Comparison to generative list-wise and point-wise rerankers}
\label{app.point.list}

\setlength{\tabcolsep}{3pt}
\begin{table}[h]
\caption{nDCG@10 on BEIR top-$40$.}
\centering
\scriptsize
\begin{tabular}{ | c | c | c  c | c c | c c | c c | }
\hline
Dataset & Retriever & \multicolumn{2}{c|}{CoRe-R} & \multicolumn{2}{c|}{MonoT5} & \multicolumn{2}{c|}{RankGPT} & \multicolumn{2}{c|}{MXBAI} \\
& baseline & (Llama) & (Qwen) & (220M) & (3B) & (GPT-3.5) & (GPT-4) & (v1) & (v2) \\
\hline\hline
Covid          & 63.1 & 75.7 & 76.4 & 74.5 & 75.8 & 74.2 & 77.0 & 75.2 & 73.0 \\
DBPedia        & 36.0 & 39.4 & 39.3 & 39.9 & 40.7 & 41.0 & 42.6 & 39.2 & 40.5 \\
SciFact        & 71.3 & 75.1 & 76.1 & 72.1 & 75.2 & 64.1 & 75.2 & 74.1 & 75.9 \\
NFCorpus       & 33.7 & 36.7 & 36.7 & 35.8 & 36.4 & 32.8 & 37.1 & 36.5 & 33.9 \\
Scidocs        & 22.5 & 20.2 & 21.6 & 15.7 & 16.6 & 15.4 & 20.1 & 19.2 & 17.7 \\
Climate        & 30.3 & 22.7 & 24.5 & 20.9 & 21.7 & 22.4 & 24.3 & 21.3 & 24.5 \\
\hline
Average        & 42.8 & 45.0 & 45.8 & 43.1 & 44.4 & 41.7 & 46.0 & 44.2 & 44.2 \\
\hline
\end{tabular}
\label{table.point}
\end{table}

We compare CoRe-R against RankGPT \cite{sun2023chatgpt}, MonoT5 \citep{nogueira2020document}, and MXBAI \cite{li2025prorank} and report the top-40 results in Table~\ref{table.point}. Even as a training-free framework, CoRe-R shows competitive performance and delivers higher averaged accuracy than MonoT5 which was specifically trained for reranking tasks. CoRe-R (Llama and Qwen) also outperforms RankGPT-3.5 and MXBAI-v1 in all tasks, and CoRe-R (Qwen) delivers competitive accuracy compared to RankGPT-4 while requiring no generation and much faster runtime.

\subsection{Results with different prompt structures}
\label{app.result.prompt}

\textbf{General prompt ablation.} We conduct additional experiment using different paraphrasing variants of the general instruction prompt as described in Appendix~\ref{app.prompt} (\texttt{"Given a list of documents: ... Answer the query: ..."}). We also evaluate the reranking tasks using an empty instruction prompt (context contains only documents and query). Table~\ref{table.prompt} shows the corresponding average nDCG@10 on BEIR using Mistral 7B.

\begin{table}[h]
    \centering
    \caption{Average nDCG@10 on BEIR using different instruction prompts with Mistral 7B.}
    \begin{tabular}{|c|cccccc|}
        \hline
        Method & original & variant 1 & variant 2 & variant 3 & variant 4 & empty \\
        \hline\hline
        ICR & 47.9 & 48.9 & 48.2 & 47.7 & 48.0 & 47.2 \\
        QR-R & 48.9 & 49.2 & 48.7 & 48.8 & 48.6 & 48.5 \\
        CoRe-R & 50.9 & 51.0 & 50.5 & 50.8 & 50.5 & 50.4 \\
        \hline
    \end{tabular}
    \label{table.prompt}
\end{table}

CoRe-R is stable across all six prompt conditions, and remains the strongest method under every prompt variant. Even when the instruction is removed entirely and the input contains only the documents and query, CoRe-R achieves 50.4, compared with 50.9 under the original prompt. These results indicate that the gains from CoRe head selection are not dependent on a narrowly optimized instruction prompt.

\textbf{Customized prompt.} All attention-based rerankers underperform the retriever baseline in some datasets that require more task-specific instruction prompts. For example, we show that a customized prompt for Quora (\texttt{Please identify question that has the same meaning with the following query}) and ArguAna (\texttt{Please find documents that contain counter-arguments to the following query}) can improve the performance.

We report the reranking results in Table~\ref{app.table.quora}. Overall, the task-specific prompt increases the nDCG@10 scores for all attention-based rerankers by large margin. As expected, CoRe-R still stands out as the best-performing reranker, delivering the best reranking accuracy across all models and outperforms the retriever baseline ($86.7$). Future work could explore different ways to optimize the task-specific prompt design for non-QA datasets, which further improve the final reranking performance.

\begin{table}[h]
    \centering
    \caption{nDCG@10 using the universal prompt and the customized prompt.}
    \label{app.table.quora}
    \begin{tabular}{| c | c | c c | c c | c c |}
        \hline
        Dataset & Method & \multicolumn{2}{c|}{Mistral 7B} & \multicolumn{2}{c|}{Llama-3.1 8B} & \multicolumn{2}{c|}{Qwen-3 4B} \\\cline{3-8}
        & & universal & customized & universal & customized & universal & customized \\
        \hline\hline
        & ICR & 81.1 & 83.2 & 78.7 & 82.9 & 71.3 & 78.9 \\
        Quora & QR-R & 71.5 & 79.9 & 74.6 & 80.8 & 61.0 & 72.7 \\
        & CoRe-R & 85.2 & 89.2 & 82.3 & 87.3 & 81.2 & 88.4 \\
        \hline
        & ICR & 45.0 & 52.8 & 42.5 & 50.2 & 39.7 & 48.9 \\
        ArguAna & QR-R & 51.1 & 54.2 & 54.2 & 56.9 & 41.4 & 51.1 \\
        & CoRe-R & 52.7 & 55.7 & 55.0 & 57.1 & 52.9 & 56.3 \\
        \hline
    \end{tabular}
    \label{tab:placeholder}
\end{table}

\subsection{Results on modern base retriever}
\label{app.qwen.retriever}

We conduct additional experiment to verify the robustness of attention-based re-rankers against more modern retriever baseline. In this test, we select Qwen3-0.6b \citep{qwen3embedding} as a moderate-size embedding model with strong retrieval capability.

\setlength{\tabcolsep}{4pt}
\begin{table}[h]
\caption{nDCG@10 on the BEIR benchmark using Llama-3.1 8B.}
\label{table.qwen}
\begin{center}
\scriptsize
\begin{tabular}{ | c | c | c c c | }
\hline
Dataset & Retriever & \multicolumn{3}{c|}{Reranker} \\\cline{3-5}
 & Qwen3-0.6b & ICR & QR-R & CoRe-R \\
\hline\hline
TREC-COVID    & 89.6 & 88.9 & 89.1 & \textbf{89.7} \\
NFCorpus      & 36.2 & 36.3 & \textbf{38.2} & \textbf{38.2} \\
DBPedia       & 39.4 & 40.1 & 41.3 & \textbf{42.0} \\
SciFact       & 69.8 & 74.7 & 75.5 & \textbf{76.3} \\
SciDocs       & 24.4 & 19.9 & 20.8 & \textbf{21.3} \\
FiQA          & 46.8 & 47.8 & 50.2 & \textbf{50.4} \\
NQ            & 53.4 & 59.0 & 60.5 & \textbf{61.7} \\
FEVER         & 87.9 & 88.1 & \textbf{88.3} & \textbf{88.3} \\
Climate-FEVER & 42.1 & 30.1 & 30.3 & \textbf{31.1} \\
HotpotQA      & 65.7 & 74.2 & 74.5 & \textbf{74.8} \\
Touche        & 32.8 & 33.4 & 33.5 & \textbf{34.9} \\
MSMARCO       & 37.8 & 36.9 & 37.0 & \textbf{37.9} \\
Quora         & 87.8 & 76.8 & 74.0 & \textbf{81.7} \\
ArguAna       & 70.9 & 45.6 & 61.5 & \textbf{62.3} \\
CQADupstack   & 45.8 & 47.1 & 47.8 & \textbf{48.7} \\
\hline
Average       & 55.4 & 53.3 & 54.9 & \textbf{56.0} \\
\hline
\end{tabular}
\end{center}
\end{table}

Table~\ref{table.qwen} reports top-$40$ re-ranking results using different attention-based re-rankers. The overall trend matches the re-ranking results in Table~\ref{table.beir}, and CoRe-R consistently outperform all baselines in most tasks. A few exception when the base retriever has higher scores includes non-QA tasks that require more task-specific instruction prompts as demonstrated in Appendix~\ref{app.result.prompt}.

\end{document}